\documentclass[12pt]{article}
\usepackage{graphicx, psfrag, epsf}
\usepackage{caption, subcaption, float, color, multirow, bbm}
\usepackage[colorlinks=true, urlcolor=blue,linkcolor=blue, citecolor=magenta, breaklinks=true]{hyperref}
\usepackage{natbib}
\usepackage{empheq}
\usepackage{enumerate} 
\usepackage{dsfont}
\usepackage{amssymb} 
\usepackage{amsthm} 
\usepackage{mathrsfs}
\usepackage{bigints} 
\usepackage{fancyhdr}

\newtheorem{thm}{Theorem}

\newtheorem{cor}[thm]{Corollary}

\newtheorem{theorem}[thm]{Theorem}
\newtheorem{assumption}[]{Assumption}

\numberwithin{equation}{section}

\newcommand{\blind}{0}

\allowdisplaybreaks

\def\boxit#1{\vbox{\hrule\hbox{\vrule\kern6pt
			\vbox{\kern6pt#1\kern6pt}\kern6pt\vrule}\hrule}}

\date{}

\begin{document}
	\sloppy
	
	\def\spacingset#1{\renewcommand{\baselinestretch}%
		{#1}\small\normalsize} \spacingset{1}

	
	\if0\blind
	{
		\title{\bf A Unified Framework for Specification Tests of Continuous Treatment Effect Models}
		\author{Wei Huang\thanks{E-mail: \texttt{wei.huang@unimelb.edu.au}}\hspace{.2cm}\\
			School of Mathematics and Statistics and Australian Research Council Centre \\of Excellence for Mathematical  and Statistical Frontiers (ACEMS),\\ University of Melbourne, Australia\\
			and\\
			Oliver Linton\thanks{E-mail: \texttt{obl20@cam.ac.uk}}\hspace{.2cm}\\
			Faculty of Economics, University of Cambridge\\
			and \\
			Zheng Zhang\thanks{\textit{Corresponding author}. E-mail: \texttt{zhengzhang@ruc.edu.cn}}\\
			Institute of Statistics \& Big Data,\\  Renmin University of China
		}
		\maketitle
	} \fi
	
	\if1\blind
	{
		\bigskip
		\bigskip
		\bigskip
		\begin{center}
			{\LARGE\bf A Unified Framework for Specification Tests of Continuous Treatment Effect Models}
		\end{center}
		\medskip
	} \fi
	
	\bigskip
	\begin{abstract}
		We propose a general framework for the specification testing of continuous
		treatment effect models. We assume a general residual function, which includes
		the average and quantile treatment effect models as special cases. The null
		models are identified under the unconfoundedness condition and contain a
		nonparametric weighting function. We propose a test statistic for the null
		model in which the weighting function is estimated by solving an expanding set
		of moment equations. We establish the asymptotic distributions of our test
		statistic under the null hypothesis and under fixed and local alternatives.
		The proposed test statistic is shown to be more efficient than that
		constructed from the true weighting function and can detect local alternatives
		deviated from the null models at the rate of $O(N^{-1/2})$. A simulation
		method is provided to approximate the null distribution of the test statistic.
		Monte-Carlo simulations show that our test exhibits a satisfactory
		finite-sample performance, and an application shows its practical value.
	\end{abstract}
	
	\noindent%
	\vfill
	
	\newpage
	\spacingset{1.45} 
	\section{Introduction}\label{sec:introduction}
	Causal inference is a central topic in economics, statistics, and machine
	learning. Although a randomized trial is the gold standard for identifying 
	causal effects, such trials are often unavailable or even unethical in practice.
	Observational data, which are collected when the participation of an intervention is only
	observed rather than manipulated by scientists, are predominantly the type of data that are 
	available. A major challenge for inferring causality in observational
	studies is confoundedness, whereby individual characteristics are
	correlated with both the treatment variable and the outcome of interest. To
	identify causality, the \emph{unconfounded treatment assignment} condition is
	frequently imposed in the literature; see 
	\cite{rosenbaum1983central,rosenbaum1984reducing}. For a comprehensive review
	of causal inference and its applications, see \cite{imbens2009recent} and
	\cite{abadie2018econometric}.
	
	Treatment effect models are used extensively in economics and statistics to
	evaluate the causal effect of a treatment or policy. Most of the existing
	literature focuses on binary treatment, whereby an individual either does or does not receive
	the treatment (e.g., \citealp{hahn1998role,Hirano03,donald2014testing,imai2014covariate,abrevaya2015estimating,chan2016globally,Athey2018Approximate,hsu2020counterfactual,chen2020stochastic,fan2020estimation,sant2018covariate,aisimple}).  Some
	studies focus on multivalued treatment (see, e.g.,
	\citealp{cattaneo2010efficient,lee2018efficient,ai2020mann,ao2021multivalued}). However, in many applications, the treatment
	variable is continuously valued, and its causal effect is of great interest to
	decision makers. For example, when evaluating how non-labor income affects the
	labor supply, the causal effect may depend on not only the introduction of the
	non-labor income but also the total non-labor income. Similarly, when evaluating
	how advertising affects the campaign contributions for political analysis, the
	causal effect may depend not only on whether any advertisements are released
	but also on how many of them are distributed.
	
	Estimation of continuous treatment effects has received considerable attention from researchers (see \citealp{hirano2004propensity,galvao2015uniformly,kennedy2017non,Fong_Hazlett_Imai_2018,dong2019regression,huber2020direct,colangelo2020double,ai2021estimation}, among others). \cite{hirano2004propensity}, \cite{galvao2015uniformly}, and \cite{Fong_Hazlett_Imai_2018} applied fully parametric methods by modeling either the conditional distribution of the treatment given the confounders or that of the observed outcome given the treatment and the confounders. The shortcoming of these parametric methods is that modeling and testing the relationship between the treatment and the observed outcome regarding the confounders are difficult, especially when multiple confounding variables are involved. If the model is mis-specified, the conclusion can be biased and completely misleading. \cite{kennedy2017non} and \cite{huber2020direct} estimated the
	continuous treatment effects using the nonparametric kernel method. Although nonparametric approaches are much more flexible than parametric ones, the former require smoothing of the data rather than estimating finite dimensional parameters, which leads to less precise fits and slower convergence rates (slower than $N^{-1/2}$). Furthermore, it is usually difficult to interpret nonparametric results.
	
	In a recent article, \cite{Ai_Linton_Motegi_Zhang_cts_treat} studied
	continuous treatment effects by imposing a \emph{univariate generalized
		parametric} model for the functionals of the potential outcome over the
	treatment variable. The general framework includes many important causal
	parameters as special cases, for example, average and quantile treatment
	effects. They proposed a generalized weighting estimator for the causal effect
	with the weights modeled nonparametrically and estimated by solving an
	expanding set of equations. They further derived the semiparametric efficiency
	bound for the causal effect of treatment under the unconfounded treatment
	assignment condition and showed that their estimator is $\sqrt{N}%
	$-asymptotically normal and attains the semiparametric efficiency bound.
	Although \cite{Ai_Linton_Motegi_Zhang_cts_treat}'s
	estimator enjoys superior asymptotic properties and satisfactory finite sample
	performance, they did not detail the specifications of the parametric models
	for the functionals of the potential outcomes. If the parametric model is
	mis-specified, the results developed in
	\cite{Ai_Linton_Motegi_Zhang_cts_treat} do not hold.
	
	We study the question of model specification. In particular, we propose a
	consistent specification test for the most generalized continuous treatment
	effect model. That is, we consider the generalized parametric model in
	\cite{Ai_Linton_Motegi_Zhang_cts_treat} as the null model while testing our hypothesis. The potential outcome variable in the model is not observable. However,
	under the unconfounded treatment assignment condition, the model can be
	identified by a semiparametric weighted conditional model. There is abundant
	literature on the specification tests for conditional models (
	e.g.,
	\cite{ait2001goodness,bierens1982consistent,bierens1990consistent,fan1996consistent,zheng1996consistent,bierens1997asymptotic,stute1997nonparametric,li1999consistent,chen1999consistent,fan2000consistent,li2003consistent,crump2008nonparametric}
	). Most authors have considered the problems of testing a
	parametric/semiparametric null model using an integrated type test
	statistic. \cite{ait2001goodness} and \cite{chen1999consistent} considered
	testing nonparametric/semiparametric null models using nonparametric
	kernel methods. \cite{li2003consistent} considered testing the
	nonparametric/semiparametric using series methods.
	\cite{crump2008nonparametric} derived a nonparametric Wald test statistic for
	testing the conditional average treatment effects under the unconfoundedness
	condition. For the binary treatment effect model, \cite{shaikh2009specification} parametrically modeled the propensity score as a conditional expectation of the treatment given the confounders and proposed an associated specification test. This differs from our problem in the sense that we consider the specification test for the function of the potential outcome with a continuous treatment variable.
	
	Specifically, we estimate our semiparametric weighted null model using the framework developed in \cite{Ai_Linton_Motegi_Zhang_cts_treat} and construct a Cramér–von Mises test statistic and a Kolmogorov–Smirnov test one to test the null model. Although the weights in our null model are estimated nonparametrically, we show that our proposed test statistic is more efficient than that constructed from the true weights. Moreover, our proposed test statistic can detect local alternatives that deviate from the null model at the rate of $O(N^{-1/2})$.
	
	Under the null hypothesis our test statistic is shown to converge in
	distribution to a weighted sum of independent chi-squared random variables. It
	is known that obtaining the exact critical values of such a distribution is
	extremely difficult in practice. Most of the literature suggests using a
	residual wild bootstrap procedure to approximate the critical values. This is
	not applicable in our case because our null model does not imply any explicit
	form of relationship among the observed outcome, the treatment, and the
	confounders for residual sampling. To resolve this problem, we  adopt a special case of the exchangeable bootstrap to approximate the null limiting distribution. Monte-Carlo
	simulations and real data analysis were conducted to demonstrate the numerical
	properties of our test method and limiting distribution approximation.
	
	The remainder of the paper is organized as follows. We introduce the problem
	formulation and notations in Section~\ref{sec:framework}. Section~\ref{sec:ts}
	constructs the test statistic, followed by the study of the asymptotic
	properties under null hypothesis, the fixed and the local alternatives in
	Section~\ref{sec:asymptotics}. In Section~\ref{sec:approximation}, we discuss
	how to approximate the limiting distribution under the null hypothesis.
	Finally, Section~\ref{sec:numerical} discusses the choice of the tuning
	parameters in the estimation and investigates the finite sample performance
	through simulations and U.S. campaign advertisement data. All
	proofs are detailed in the supplementary file.

	\section{Basic framework}
	\label{sec:framework}
	Let $T$ denote a continuous treatment variable with
	support $\mathcal{T}\subset\mathbb{R}$, where $\mathcal{T}$ is a continuum
	subset, and $T$ has a marginal density function $f_{T}(t)$. Let $Y^{\ast}(t)$
	denote the potential response when treatment $T=t$ is assigned. We are
	interested in testing the null hypothesis:
	\begin{align}
	\label{null_1}H_{0}: \exists\ \text{some}\ \boldsymbol{\theta}^{*}\in
	\Theta,\ \text{s.t.}\ \mathbb{E}[m\{Y^{*}(t);g(t;\boldsymbol{\theta}%
	^{*})\}]=0\ \text{for all $t\in\mathcal{T}$}\,,
	\end{align}
	against the alternative hypothesis
	\begin{align*}
	H_{1}: \nexists\ \text{any}\ \boldsymbol{\theta}\in\Theta,\ \text{s.t.}%
	\ \mathbb{E}[m\{Y^{*}(t);g(t;\boldsymbol{\theta})\}]=0\ \text{for all
		$t\in\mathcal{T}$},
	\end{align*}
	where $\Theta$ is a compact set in $\mathbb{R}^{p}$ for some integer $p\geq1$,
	$m(\cdot)$ is some generalized residual function which could possibly be
	\emph{non-differentiable}, and $g(t;\boldsymbol{\theta})$ is a parametric
	working model which is differentiable with respect to $\boldsymbol{\theta}$. If $H_{0}$ holds, for each $t$, the dose-response function (DRF)
	is defined as the value $g(t;\boldsymbol{\theta}^{*})$ that solves the moment
	condition in \eqref{null_1}. The following examples show that the average
	dose-response function (ADRF) and the quantile
	dose-response function (QDRF) are special cases of
	$g(t;\boldsymbol{\theta}^{*})$, which result from choosing specific forms of $m(\cdot)$.
	\begin{itemize}
		\item(Average)  Setting  $m\left\{ Y^{*}(t);g(t;\boldsymbol{\theta}%
		^{*})\right\}  = Y^{*}(t)- g(t;\boldsymbol{\theta}^{*})$ and letting its first
		moment equal zero for each $t$, we obtain $g(t;\boldsymbol{\theta}%
		^{*})=\mathbb{E}\{Y^{*}(t)\}$, the unconditional  ADRF, which is also called
		the  \emph{marginal structural model} \citep{robins2000marginal} and the \emph{average structural function} in nonseparable models \citep{blundell2001endogeneity,imbens2009identification}. This can
		recover the average treatment effect (ATE), which is  given by $\text{ATE}%
		(t_{1},t_{0})=\mathbb{E}\{Y^{*}(t_{1})\}-\mathbb{E}\{Y^{*}(t_{0})\}$.
		Examples  include the linear marginal structure model $\mathbb{E}\{Y^{\ast
		}(t)\}=\beta_{0}+\beta_{1}\cdot t$, and the nonlinear marginal structure
		model  $\mathbb{E}\{Y^{\ast}(t)\}=\beta_{0}\cdot t+1/(t+\beta_{1})^{2}$
		studied in  \cite{hirano2004propensity}). 
		\item (Quantile) Let $\tau\in(0,1)$ and $F_{Y^{*}(t)}(\cdot)$ be the
		cumulative distribution function of $Y^{*}(t)$. Setting  $m\left\{
		Y^{*}(t);g(t;\boldsymbol{\theta}^{*})\right\}  =\tau- \mathbbm{1}\{Y^{*}%
		(t)<g(t;\boldsymbol{\theta}^{*})\}$ and letting its first moment  equal zero
		for each $t$, we obtain $g(t;\boldsymbol{\theta}^{*})=F^{-1}_{Y^{*}(t)}%
		(\tau):=\inf\{ q:\mathbb{P}(Y^{\ast}(t)\geq q)\leq\tau\} $, the unconditional
		QDRF, which is also called
		the  \emph{quantile structural model} \citep{imbens2009identification}. This can recover the quantile treatment effect (QTE), which is  given by
		$\text{QTE}(t_{1},t_{0})=F^{-1}_{Y^{*}(t_{1})}(\tau)-F^{-1}_{Y^{*}(t_{0}%
			)}(\tau)$. See \cite{Firpo2007Efficient} for detailed discussion on QTE.
		Examples include the linear model $g(t;\boldsymbol{\theta}) =\theta_{0}%
		+\theta_{1}\cdot t$ and the Box-Cox  transformation model
		$g(t;\boldsymbol{\theta}) =h_{\lambda}\left(  \theta_{0}+\theta_{1}\cdot
		t\right)  $ studied  in \cite{buchinsky1995quantile}, where $h_{\lambda
		}(z)=(\lambda z+1)^{-1/\lambda}$.
	\end{itemize}
	
	We consider an observational study in which the potential outcome $Y^{\ast }(t)$ is not observed for all $t$. Let $Y:=Y^{\ast }(T)$ denote the observed
	response. Under the null hypothesis, one may attempt to solve the following equation to find $\boldsymbol{\theta}^*$:
	\begin{equation*}
	\mathbb{E}[m\{Y;g(T;\boldsymbol{\theta})\}|T]=0.
	\end{equation*}
	However, if there is a selection into treatment, even under the null hypothesis, the true value $\boldsymbol{\theta }^*$ does not solve the above equation.
	Indeed, in this case, the observed response and the treatment assignment data
	alone cannot identify ${\boldsymbol{\theta}}^*$. To address this
	identification issue, most studies in the literature impose a selection on
	the observable condition \citep[e.g.,][]{Hirano03, imai2004causal, Fong_Hazlett_Imai_2018,Ai_Linton_Motegi_Zhang_cts_treat}.
	Specifically, let $\boldsymbol{X}\in\mathbb{R}^r$, for some integer $r\geq 1$, denote a vector of observable covariates. The following condition shall be maintained throughout the paper.
	
	\begin{assumption}[\emph{Unconfounded Treatment Assignment}]
		\label{as:TYindep} For all $t\in \mathcal{T}$, given $\boldsymbol{X}$, $T$ 
		is independent of $Y^{\ast }(t)$, that is, $Y^{\ast }(t)\perp T|\boldsymbol{X}$,
		for all $t\in \mathcal{T}$.
	\end{assumption}
	
	Let $\{T_i,\boldsymbol{X}_i,Y_i\}_{i=1}^N$ be an independent and identically distributed ($i.i.d.$) sample drawn from the joint distribution of $(T,\boldsymbol{X},Y)$. Let $f_{T|X}$ denote the conditional density of $T$ given
	the observed covariates $\boldsymbol{X}$. Under Assumption~\ref{as:TYindep}, \cite{Ai_Linton_Motegi_Zhang_cts_treat} showed that $\mathbb{E}[m\{Y^*(t);g(t;\boldsymbol{\theta})\}]$ can be identified as follows: 
	\begin{equation*}
	\mathbb{E}[m\{Y^*(t);g(t;\boldsymbol{\theta})\}]=\mathbb{E}[\pi_0(T,\boldsymbol{X})m\{Y;g(T;\boldsymbol{\theta})\}|T=t], \ \forall t\in\mathcal{T}\,
	\end{equation*}
	where
	\begin{equation*}
	\pi _{0}(T,\boldsymbol{X}):=\frac{f_{T}(T)}{f_{T|X}(T|\boldsymbol{\ X})}\,.
	\end{equation*}
	The function $\pi_{0}(T,\boldsymbol{X})$ is called the \emph{stabilized weights} in \cite{robins2000marginal}. 
	
	The null and alternative hypothesis in \eqref{null_1} can then be re-written as
	\begin{equation}\label{CIA}
	H_0: \mathbb{P}\left(\mathbb{E}[\pi_0(T,\boldsymbol{X})m\{Y;g(T;\boldsymbol{\theta}^*)\}|T]=0\right)=1\ \text{for some $\boldsymbol{\theta}^*\in \Theta$}\,,
	\end{equation}
	against the alternative hypothesis
	\begin{align*}
	H_1: \mathbb{P}\left(\mathbb{E}[\pi_0(T,\boldsymbol{X})m\{Y;g(T;\boldsymbol{\theta})\}|T]\neq 0\right)>0\ \text{for all $\boldsymbol{\theta}\in\Theta$}\,.
	\end{align*}
	This converts the test for \eqref{null_1} to a specification test for a univariate regression model, if both $\pi_0(T,\boldsymbol{X})$ and $\boldsymbol{\theta}^*$ were given. Specially, letting 
	\begin{equation}
	U_i:=\pi_0(T_i,\boldsymbol{X}_i)m\{Y_i;g(T_i;\boldsymbol{\theta}^*)\}\,, \label{Udef}
	\end{equation}
	the null hypothesis $H_0$ is equivalent to $\mathbb{P}\{\mathbb{E}(U_i|T_i)=0\}=1$.
	A popular technique for testing such a conditional moment model is to convert it to an unconditional one. 
	
	Note that $\mathbb{P}\{\mathbb{E}(U_i|T_i)=0\}=1$ if and only if $\mathbb{E}\{U_iM(T_i)\}=0$ for all bounded and measurable functions $M(\cdot)$. Following
	\cite{bierens1997asymptotic}, \cite{stinchcombe1998consistent}, \cite{stute1997nonparametric}, and \cite{li2003consistent}, by choosing a proper weight function $\mathscr{H}(\cdot,\cdot)$, $\mathbb{E}(U_i|T_i)=0$ is a.s. equivalent to
	\begin{align}\label{eq:test_unconditional}
	\mathbb{E}\left\{U_i\mathscr{H}(T_i,t)\right\}=0 \ \text{for all}\ t\in \mathcal{T}\,.
	\end{align}
	Popular choices of such a weight function are the logistic function $\mathscr{H}(T_i,t)=1/\{1+\exp(c-t\cdot T_i)\}$ with $c\neq 0$, cosine-sine function $\mathscr{H}(T_i,t)=\cos(t\cdot T_i)+\sin(t\cdot T_i)$ and the indicator function $\mathscr{H}(T_i,t)=\mathds{1}(T_i\leq t)$ (see \citealp{stinchcombe1998consistent} and \citealp{stute1997nonparametric} for more detailed discussion). Now, letting
	\begin{align}
	J_N^0(t)=\frac{1}{\sqrt{N}}\sum_{i=1}^NU_i\mathscr{H}(T_i,t)\,,\label{J0N}
	\end{align}
	the sample analogue of $\mathbb{E}\left\{U_i\mathscr{H}(T_i,t)\right\}$ multiplied by $\sqrt{N}$, one can test $H_0$ using the Cramér–von Mises (CM)-type statistic
	\begin{align}
	CM^0_N=\int \{J^0_N(t)\}^2\widehat{F}_T(dt)=\frac{1}{N}\sum_{i=1}^N\{J^0_N(T_i)\}^2,\label{CM0N}
	\end{align}
	or the Kolmogorov-Smirnov (KS)-type statistic
	\begin{align}\label{KS0N}
	KS_N^0=\sup_{t\in\mathcal{T}}|J^0_N(t)|,
	\end{align}
	where $\widehat{F}_T(\cdot)$ is the empirical distribution of $T_1,...,T_N$. 
	However, both $\pi_0(T,\boldsymbol{X})$ and $\boldsymbol{\theta}^*$ are unknown in practice so that the $U_i$'s are unavailable. We must replace the $U_i$'s with some estimates, which is studied in the following section.
	
	\noindent \emph{\textbf{Remark 1}. Note that in our model, the stabilized weights $\pi_{0}(T,\boldsymbol{X})$ are nonparametric. The conditional distribution of the treatment, given the confounders $f_{T|X}(T|\boldsymbol{X})$, which is known as the \emph{generalized propensity score} \citep{hirano2004propensity}, is also nonparametric. Under the unconfoundedness assumption, an alternative identification  of the null model is through the conditional distribution of the outcome given the treatment and the confounders and the marginal distribution of the confounders, that is, $\mathbb{E}[m\{Y^*(t);g(t;\boldsymbol{\theta})\}]=\mathbb{E}\Big(\mathbb{E}[m\{Y;g(T;\boldsymbol{\theta})\}|\boldsymbol{X},T=t]\Big)$ holds under Assumption \ref{as:TYindep}.} 
	
	\section{Test statistic}\label{sec:ts}
	One obvious approach for estimating the $U_i$'s is to estimate $f_{T}(T_i)$ and $f_{T|X}(T_i|\boldsymbol{X}_i)$, then construct the estimators of $\pi_0(T_i,\boldsymbol{X}_i)$ and $\boldsymbol{\theta}^*$. However, it is well-known that this ratio estimator of $\pi_0(T,\boldsymbol{X})$ is very sensitive to small values of $
	f_{T|X}(T|\boldsymbol{X})$ because small estimation errors in estimating $
	f_{T|X}(T|\boldsymbol{X})$ result in large
	estimation errors of the estimator of $\pi_0(T,\boldsymbol{X})$. To avoid or mitigate this problem, \cite{Ai_Linton_Motegi_Zhang_cts_treat} directly estimated $\pi_{0}(T,\boldsymbol{X})$ as a whole using the generalized empirical likelihood (GEL). We adopt their estimator and elaborate its construction as follows. Note that the weighting function satisfies  
	\begin{equation}
	\mathbb{E}\left\{\pi_{0}(T,\boldsymbol{X})u(T)v(\boldsymbol{X})\right\} =%
	\mathbb{E}\{u(T)\}\cdot\mathbb{E}\{v(\boldsymbol{X})\}\   \label{moment1}
	\end{equation}
	for any suitable functions $u(t)$ and $v(\boldsymbol{x})$. \citet[Theorem 2]{Ai_Linton_Motegi_Zhang_cts_treat} showed that the restriction \eqref{moment1} identifies the weighting function $\pi_{0}(T,\boldsymbol{X})$. This result suggests that one may estimate the $\pi_0(T_i,\boldsymbol{X}_i)$'s by solving the sample analogue of \eqref{moment1}. The challenge is that (\ref{moment1}) implies an infinite number of equations, which is impossible to solve with a finite sample of observations. To overcome this difficulty, \cite{Ai_Linton_Motegi_Zhang_cts_treat} suggested approximating
	the infinite-dimensional function space by a sequence of finite-dimensional sieve spaces. Specifically, let $u_{K_{1}}(T)=(u_{K_{1},1}(T),%
	\ldots,u_{K_{1},K_{1}}(T))^{\top }$ and $v_{K_{2}}(\boldsymbol{X}%
	)=\big( v_{K_{2},1}(\boldsymbol{X}),\ldots,$ $v_{K_{2},K_{2}}(\boldsymbol{X}%
	)\big) ^{\top }$ denote some known basis functions with dimensions $%
	K_{1}\in \mathbb{\ N}$ and $K_{2}\in \mathbb{N}$ respectively, and let $%
	K:=K_{1}\cdot K_{2}$. The functions $u_{K_{1}}(t)$ and $v_{K_{2}}(%
	\boldsymbol{x})$ are called the \emph{approximation sieves}, such as B-splines or power series \citep[see][for more discussion on sieve approximation]{Newey97, chen2007large}. Because the sieve approximating space is a subspace of the original function space, $%
	\pi_{0}(T,\boldsymbol{X})$ also satisfies  
	\begin{equation}
	\mathbb{E}\left\{\pi_{0}(T,\boldsymbol{X})u_{K_{1}}(T)v_{K_{2}} (\boldsymbol{X})^{\top}\right \}=\mathbb{E}\{u_{K_{1}}(T)\}\cdot \mathbb{E}\{v_{K_{2}}(%
	\boldsymbol{X})\}^{\top}\,.   \label{sievemoment}
	\end{equation}
	
	Following \cite{Ai_Linton_Motegi_Zhang_cts_treat}, we estimate the $\pi_0(T_i,\boldsymbol{X}_i)$'s consistently by the $\widehat{\pi}_i$'s that maximize the generalized empirical likelihood (GEL) function, subject to the sample analog of \eqref{sievemoment}:
	{\small \  
		\begin{equation}
		\left\{ 
		\begin{array}{cc}
		& \left\{\widehat{\pi}_{i}\right\} _{i=1}^{N}=\arg\max\left( -N^{-1}
		\sum_{i=1}^{N}\pi_{i}\log\pi_{i}\right) \\[2mm] 
		& \text{subject to}\ \frac{1}{N}\sum_{i=1}^{N}%
		\pi_{i}u_{K_{1}}(T_{i})v_{K_{2} }(\boldsymbol{X}_{i})^{\top}=\left\{ \frac{1}{%
			N}\sum_{i=1}^{N}u_{K_{1}} (T_{i})\right\} \left\{\frac{1}{N}%
		\sum_{j=1}^{N}v_{K_{2}}(\boldsymbol{X} _{j})^{\top}\right\}\,.%
		\end{array}
		\right.   \label{E:cm1}
		\end{equation}}Two observations are immediately clear. First, by including a constant of one in
	the sieve base functions, \eqref{E:cm1} guarantees that $N^{-1}\sum_{i=1}
	^{N}\widehat{\pi}_{i}=1$. Second, we notice that  
	\begin{equation*}
	\max\left( -N^{-1}\sum_{i=1}^{N}\pi_{i}\log\pi_{i}\right) =-\min\left\{ 
	\sum_{i=1}^{N}(N^{-1}\pi_{i})\cdot\log\left(\frac{N^{-1}\pi_{i}}{N^{-1}}\right)\right\}\,.
	\end{equation*}
	The entropy maximization problem minimizes the Kullback-Leibler divergence 
	between the weights $\{N^{-1}\pi_{i}\}_{i=1}^{N}$ and the empirical 
	frequencies $\{N^{-1}\}$, subject to the sample analogue of  %
	\eqref{sievemoment}. Further, \cite{Ai_Linton_Motegi_Zhang_cts_treat} showed that the dual solution of the primal problem \eqref{E:cm1} is
	\begin{align}  \label{def:pihat_dual}
	\widehat{\pi}_{K}(T_{i},\boldsymbol{X}_{i}):=\rho ^{\prime }\left\{
	u_{K_{1}}(T_{i})^{\top }\widehat{\Lambda}_{K_{1}\times K_{2}}v_{K_{2}}(%
	\boldsymbol{X}_{i})\right\} ,
	\end{align}%
	where $\rho^\prime$ is the first derivative of $\rho$ with $\rho(u)=-\exp(-u-1)$, and $\widehat{\Lambda}_{K_{1}\times K_{2}}$ is the maximizer of the strictly
	concave function $\widehat{G}_{K_{1}\times K_{2}}$ defined by
	\begin{align}
	&\widehat{G}_{K_{1}\times K_{2}}(\Lambda )\nonumber\\
	&:=\frac{1}{N}\sum_{i=1}^{N}\rho \left\{u_{K_{1}}(T_{i})^{\top }\Lambda v_{K_{2}}(\boldsymbol{X}_{i})\right\} -\left\{\frac{1}{N}\sum_{i=1}^{N}u_{K_{1}}(T_{i})\right\} ^{\top }\Lambda \left\{\frac{1}{N}\sum_{j=1}^{N}v_{K_{2}}(\boldsymbol{X}_{j})\right\}.\label{def:G^hat}
	\end{align}
	The first order condition of \eqref{def:G^hat} implies that $\{\widehat{\pi}_{K}(T_{i},\boldsymbol{X}_{i})\}_{i=1}^N$ satisfies the sample analog of \eqref{sievemoment}; such restrictions reduce the chance of obtaining extreme weights. The concavity of \eqref{def:G^hat} enables us to obtain the solution quickly via the Gauss-Newton algorithm. To ensure a consistent estimate of $\pi_0(T,\boldsymbol{X})$, the dimensions of the bases, $K_1$ and $K_2$, shall increase as the sample size increases. The choice of $K_1$ and $K_2$ in practice will be discussed in Section~\ref{CV}.
	
	 Having estimated the weights, we now propose an extremum estimator for  $\boldsymbol{\theta}^*$ (e.g.,~\citealp{pakes1989simulation}, \citealp{chen2003estimation}, \citealp{de2019smoothed}). Note that under $H_0$, the true value $\boldsymbol{\theta}^*$ solves the following equation: 
	\begin{equation}\label{newmodel}
	\mathbb{E}\left[\pi_0(T,\boldsymbol{X})m\{Y;g(T;\boldsymbol{\theta})\}w(T;\boldsymbol{\theta})\right]=0, 
	\end{equation}
	where $w(T;\boldsymbol{\theta})$ (which may  possibly not involve $\boldsymbol{\theta}$) is a prespecified $q$-dimensional vector with $q\geq p$ such that, under $H_0$, $\boldsymbol{\theta}^*$ is identified or over-identified. Examples of such vectors include $w(T;\boldsymbol{\theta})=(1,T,...,T^{q-1})^\top$ or $w(T;\boldsymbol{\theta})=\nabla_{\boldsymbol{\theta}}g(T;\boldsymbol{\theta})$, where $``\nabla_{\boldsymbol{\theta}}"$ denotes the derivative with respect to $\boldsymbol{\theta}$. 
	We then estimate $\boldsymbol{\theta }^*$ by
	\begin{align}\label{def:thetahat}
	\widehat{\boldsymbol{\theta }}:=\arg\min_{\boldsymbol{\theta}\in\Theta}\left\|M_N(\boldsymbol{\theta},\widehat{\pi}_K)\right\|,
	\end{align}	
	where $\|\cdot\|$ is the Euclidean norm, and
	\begin{align*}
	M_N(\boldsymbol{\theta},\pi):=\frac{1}{N}\sum_{i=1}^{N}\pi(T_{i},
	\boldsymbol{X}_{i})m\{Y_{i};g(T_{i};\boldsymbol{\theta})\}w(T_{i}; \boldsymbol{\theta})\,.
	\end{align*}
	With the estimators $\{\widehat{\pi}_K(T_i,\boldsymbol{X}_i)\}_{i=1}^N$ of $\{\pi_0(T_i,\boldsymbol{X}_i)\}_{i=1}^N$ and $\widehat{\boldsymbol{\theta}}$ of $\boldsymbol{\theta}$, we estimate $U_i$ by $ \widehat{U}_i=\widehat{\pi}_{K}(T_i,\boldsymbol{X}_i)m\{Y_i;g(T_i;\widehat{\boldsymbol{\theta }})\}$, for $i=1,\ldots,N$. Replacing the $U_i$'s in \eqref{J0N} by the $\widehat{U}_i$'s, we have a feasible test statistic for $H_0$  based on
	\begin{align*}
	\widehat{J}_N(t)=\frac{1}{\sqrt{N}}\sum_{i=1}^N\widehat{U}_i\mathscr{H}(T_i,t)\,, 
	\end{align*}
	the corresponding estimators of the Cramér–von Mises (CM)-type statistic in \eqref{CM0N} and the Kolmogorov-Smirnov (KS)-type statistic in \eqref{KS0N} are, respectively,
	\begin{align}\label{test:CMandKS}
	\widehat{CM}_N=\frac{1}{N}\sum_{i=1}^N\{\widehat{J}_N(T_i)\}^2 \quad\text{and}\quad \widehat{KS}_N=\sup_{t\in\mathcal{T}}|\widehat{J}_N(t)|\,,
	\end{align}
	where the supremum is calculated as the maximum value over a discretization of $\mathcal{T}$ in practice.
	
	\noindent \emph{\textbf{Remark 2}.  An alternative estimator of
		$\boldsymbol{\theta}^{\ast}$ can be constructed under $H_{0}$. Suppose that,
		under $H_{0}$, $\boldsymbol{\theta}^{\ast}$ is identified by the unique
		solution to the following optimization problem:
		\[
		\boldsymbol{\theta}^{\ast}=\arg\min_{\boldsymbol{\theta}\in\Theta
		}CM(\boldsymbol{\theta}):=N\times\int_{\mathcal{T}}\left\{  \mathbb{E}\left[
		U_{i}(\boldsymbol{\theta})\mathcal{H}(T_{i},t)\right]  \right\}  ^{2}%
		f_{T}(t)dt,
		\]
		where $U_{i}(\boldsymbol{\theta}):=\pi_{0}(T_{i},\boldsymbol{X}_{i}%
		)m\{Y_{i};g(T_{i};\boldsymbol{\theta})\}$. Let $\widehat{U}_{i}%
		(\boldsymbol{\theta}):=\widehat{\pi}_{K}(T_{i},\boldsymbol{X}_{i}%
		)m\{Y_{i};g(T_{i};\boldsymbol{\theta})\}$ and $\widehat{J}_{N}%
		(t;\boldsymbol{\theta}):=N^{-1/2}\sum_{i=1}^{N}\widehat{U}_{i}%
		(\boldsymbol{\theta})\mathscr{H}(T_{i},t)$. Under $H_{0}$, the estimator of
		$\boldsymbol{\theta}^{\ast}$ can be defined by
		\begin{align}
		\widehat{\boldsymbol{\theta}}_{opt}:=\arg\min_{\boldsymbol{\theta}\in\Theta
		}\widehat{CM}_{N}(\boldsymbol{\theta}):=\arg\min_{\boldsymbol{\theta}\in
		\Theta}\frac{1}{N}\sum_{i=1}^{N}\{\widehat{J}_{N}(T_{i};\boldsymbol{\theta
	})\}^{2}.
	\end{align}
	Therefore, the alternative test statistic is $\widehat{CM}_{N}%
	(\widehat{\boldsymbol{\theta}}_{opt})$. However, seeking the global minimizer
	of $\widehat{CM}_{N}(\boldsymbol{\theta})$ is difficult as $\widehat{CM}%
	_{N}(\boldsymbol{\theta})$ may not be differentiable, convex, and even 
	continuous. For example, taking $m\{Y_{i};g(T_{i};\boldsymbol{\theta}%
	)\}=\tau-\mathds{1}\{Y_{i}\leq g(T_{i};\boldsymbol{\theta})\}$ for QDRF,  a unique solution to the problem does not exist. Under a stronger
	condition that $m(y;g)$ is differentiable in $g$, we establish the asymptotic
	results for both $\widehat{J}_{N}(t;\widehat{\boldsymbol{\theta}}_{opt})$ and
	$\widehat{CM}_{N}(\widehat{\boldsymbol{\theta}}_{opt})$ in section~E in the supplementary file.\\
	\noindent\textbf{Remark 3.}	 In order to estimate $\pi_0(T,\boldsymbol{X})$, \cite{Fong_Hazlett_Imai_2018} noted the moment conditions%
	\begin{align}\label{eq:fong}
	\mathbb{E}\left[  \pi_0(T,\boldsymbol{X})T\boldsymbol{X}\right] = \mathbb{E}(T)\mathbb{E}(\boldsymbol{X}), \quad \mathbb{E}\left[  \pi_0(T,\boldsymbol{X})T\right] = \mathbb{E}(T), \quad \mathbb{E}\left[  \pi_0(T,\boldsymbol{X})\boldsymbol{X}\right] = \mathbb{E}(\boldsymbol{X})\,,
	\end{align}
	which are special cases of our moment condition~\eqref{sievemoment}.
	They then proposed estimating $\pi_0(T,\boldsymbol{X})$ by maximizing the empirical likelihood of $T$ and $\boldsymbol{X}$ under the constraints of the sample analogue of \eqref{eq:fong} and estimating $\mathbb{E}\{Y^\ast(t)\}$ by a simple linear model. This can be considered as fixing $u_{K_1}(T)=(1,T)^\top$ and $v_{K_2}(\boldsymbol{X})=(1,\boldsymbol{X}^\top)^\top$, taking $m\{Y^\ast(t),g(t,\boldsymbol{\theta}^\ast)\}=Y^\ast(t) - g(t,\boldsymbol{\theta}^\ast)$ and $g$ as a simple linear model in the estimation method of \cite{Ai_Linton_Motegi_Zhang_cts_treat}. 
	However, the equation \eqref{eq:fong} is of finite dimension and cannot nonparametrically identify $\pi_0(T,\boldsymbol{X})$. Hence, \cite{Fong_Hazlett_Imai_2018} imposed a
	parametric model for the stabilized weights to achieve consistent estimation. We adopt the estimator proposed by \cite{Ai_Linton_Motegi_Zhang_cts_treat} that does not impose any parametric structure on the stabilized weights.\\
	\noindent\textbf{Remark 4}. Once our specification test rejects the null model and no better parametric model can be proposed, several solutions are available. For example, one may consider the double robustness estimator. \cite{colangelo2020double} estimate the average dose-response function $\mathbb{E}[Y^*(t)]$ based on the following double robustness representation: 
	\begin{align*}
	\mathbb{E}[Y^*(t)]=\mathbb{E}\left\{\mathbb{E}\left[Y|T=t,\boldsymbol{X}\right]+\lim_{h\to 0}\mathbb{E}\left[\frac{K_h\left(T-t\right)}{f_{T|X}(t|\boldsymbol{X})}\left\{Y-\mathbb{E}\left[Y|T=t,\boldsymbol{X}\right]\right\}\right]\right\},
	\end{align*}
	where $K_h\left(T-t\right)$ is a kernel weighting observation $T$ with treatment value of approximately $t$ in a distance of $h$. They estimate both the general propensity score $f_{T|X}(t|\boldsymbol{X})$ and the outcome regression function $\mathbb{E}\left[Y|T=t,\boldsymbol{X}\right]$ using nonparametric techniques with cross-fitting. \\
	An alternative solution to the misspecification of the dose-response function $g(t;\boldsymbol{\theta})$ is to consider a fully nonparametric specification $g(t)$, that is, to estimate $g(t)$ from the moment  $\mathbb{E}[m(Y^*(t);g(t))]=0$ for all $t\in\mathcal{T}$. Under Assumption \ref{as:TYindep}, $g(t)$ can be identified through the conditional moment $\mathbb{E}[\pi_0(T,\boldsymbol{X})m(Y;g(T))|T]=0$. We can define the sieve minimum distance (SMD)  estimator \citep{ai2003efficient} of $g(T)$ by
	\begin{align*}
	\widehat{g}(\cdot):=\arg\min_{h(\cdot)\in \mathcal{H}_{K_3}} \frac{1}{N}\sum_{i=1}^N \left\{\widehat{E}[\widehat{\pi}_K(T,\boldsymbol{X})m(Y;h(T))|T=T_i]\right\}^2\Sigma^{-1}(T_i),
	\end{align*}
	where $\Sigma(T_i)$ is a user-specified weighting function, and
	\begin{align*}
	\widehat{E}[\widehat{\pi}_K(T,\boldsymbol{X})m(Y;h(T))|T]:=&\left[\sum_{i=1}^N\widehat{\pi}_K(T_i,\boldsymbol{X})m(Y_i;h(T_i))u^\top_{K_3}(T_i)\right]\\
	&\times \left[\sum_{i=1}^Nu_{K_3}(T_i)u^\top_{K_3}(T_i)\right]^{-1}u_{K_3}(T),
	\end{align*}
	and $\mathcal{H}_{K_3}:=\{\lambda^{\top} u_{K_3}(T):\lambda\in\mathbb{R}^{K_3}\}$ is a linear sieve space. \citet[Theorem 6]{Ai_Linton_Motegi_Zhang_cts_treat} established the large sample property for the nonparametric estimator of the average dose-response function $\widehat{E}[\widehat{\pi}_K(T,\boldsymbol{X})Y|T=t]$. \\
	The extension of these methods to the general dose-response function including the quantile dose-response and the development of the corresponding large sample property is beyond the scope of this paper. \\
	\noindent\textbf{Remark 5}.\label{Remark5} If the residual function $m(y;g(t;\boldsymbol{\theta}))$ is smooth in $(t,y)$, the sieve minimum distance (SMD) estimator of $\boldsymbol{\theta}^*$ developed by \cite{ai2003efficient} is semiparametrically efficient with respect to the conditional  model $\mathbb{E}[\pi_0(T,\boldsymbol{X})m\{T;g(T;\boldsymbol{\theta}^*)\}|T]=0$.  This efficient estimation result can also be achieved based on our unconditional moment \eqref{newmodel}; indeed, by replacing $\pi_0(T,\boldsymbol{X})$ with its estimate $\widehat{\pi}_K(T,\boldsymbol{X})$ and setting $w(T;\boldsymbol{\theta})$ to be a $q$-dimensional sieve basis, for example,  $w(T;\boldsymbol{\theta})=(1,T,...,T^{q-1})^\top$, with $q\to \infty$  at an appropriate rate, it can be shown that the generalized method of moments (GMM) \citep{hansen1982large} estimator of $\boldsymbol{\theta}^*$ constructed from \eqref{newmodel} is asymptotically equivalent to the SMD estimator (see \cite{ai2020simple} for an analogous finding). However, if the residual function  $m(y;g(t;\boldsymbol{\theta}))$ is non-smooth, it remains an open problem regarding whether the efficient estimation of $\boldsymbol{\theta}^*$ from the conditional model $\mathbb{E}[\pi_0(T,\boldsymbol{X})m\{T;g(T;\boldsymbol{\theta}^*)\}|T]=0$ can be established. For this reason, and to avoid introducing an extra tuning parameter $q$, we  estimate $\boldsymbol{\theta}^*$ through \eqref{newmodel} with $w(T;\boldsymbol{\theta})$ as a fixed vector.
}

\section{Large sample properties}\label{sec:asymptotics}
This section studies the asymptotic properties of $\widehat{J}_N(\cdot)$, the test statistics $\widehat{CM}_N$ and $\widehat{KS}_N$.

\subsection{Asymptotic properties under null hypothesis}
To establish the asymptotic properties of $\widehat{J}_N(\cdot)$, $\widehat{CM}_N$ and $\widehat{KS}_N$, the following additional assumptions are imposed.

\begin{assumption}\label{as:first_order} Under $H_0$, 
	(i)  $\boldsymbol{\theta}^*$ is an interior point of $\Theta$, where $\Theta$ is a compact set in $\mathbb{R}^p$; (ii)  
	$\|M_N(\widehat{\boldsymbol{\theta}},\widehat{\pi}_K)\|=\inf_{\boldsymbol{\theta}\in\Theta_{\delta}}\left\|M_N(\boldsymbol{\theta},\widehat{\pi}_K)\right\|+o_P(N^{-1/2})$, where $\Theta_{\delta}:=\{\boldsymbol{\theta}\in\Theta:\|\boldsymbol{\theta}-\boldsymbol{\theta}^*\|\leq\delta \}$.
\end{assumption}
\begin{assumption}\label{as:var_Y}
	Let $\eta(T,\boldsymbol{X},Y;t)$ be defined in  \eqref{def:eta},	 $Var\{\eta(T,\boldsymbol{X},Y;t)\}<\infty$ for all $t\in\mathcal{T}$.
\end{assumption}

\begin{assumption}
	\ \label{as:m_smooth}
	\begin{enumerate}
		\item[(i)]  $g(t;\boldsymbol{\theta})$ is twice continuously differentiable in $
		\boldsymbol{\theta}\in\Theta$;
		\item[(ii)] $\mathbb{E}[m\{Y;g(T;\boldsymbol{\theta}^{*})\}|T=t, \boldsymbol{X}=
		\boldsymbol{x}]$ is continuously differentiable in $(t, \boldsymbol{x})$;
		\item[(iii)] $\mathbb{E}\left[\pi_{0}(T,\boldsymbol{X})m\{Y;g(T;\boldsymbol{\theta})\}w(T;
		\boldsymbol{\theta})|T=t, \boldsymbol{X}=
		\boldsymbol{x}\right] $ is differentiable w.r.t. $\boldsymbol{\theta}$ and $\nabla_{\boldsymbol{\theta}}\mathbb{E}[
		\pi_{0}(T,\boldsymbol{X})m\{Y;g(T;\boldsymbol{\theta})\}w(T;
		\boldsymbol{\theta})]  \big|_{\boldsymbol{\theta}=\boldsymbol{\theta}
			^{*}}$ is of full (column) rank.\label{Assump4iii}
	\end{enumerate}
\end{assumption}

\begin{assumption}
	\label{as:entropy} (i) $\mathbb{E}\left[\sup_{\boldsymbol{\theta} \in\Theta
	}|m\{Y;g(T;\boldsymbol{\theta})\}|^{2+\delta}\right]  <\infty$ for
	some $\delta>0$; (ii) The function class $\Big\{m\{Y;g(T;\boldsymbol{\theta})\}:\boldsymbol{\theta} \in\Theta\Big\}$ satisfies:
	\[
	\mathbb{E}\left[\sup_{\boldsymbol{\theta} _{1}:\Vert\boldsymbol{\theta}_{1}-
		\boldsymbol{\theta} \Vert<\delta}\left\vert m\{Y;g(T;\boldsymbol{\theta} _{1})\}-m\{Y;g(T;\boldsymbol{\theta})\}\right\vert ^{2}\right]  ^{1/2}\leq C\cdot\delta
	\]
	for any $\boldsymbol{\theta}\in\Theta$ and any small $\delta>0$ and for some finite positive constant $C$.
\end{assumption}
Assumption \ref{as:first_order} is
essentially stating that the estimating equation is a.s. approximately
satisfied; see \cite{pakes1989simulation} and \cite{chen2003estimation}. Assumption \ref{as:var_Y} is needed to bound the asymptotic variance of the test statistic. 
Assumption \ref{as:m_smooth} (i) and (ii) impose sufficient regularity conditions on both the link function $g$ and residual function $m$. Assumption \ref{as:m_smooth} (iii) ensures that the variance of the test statistic is finite. Assumption \ref{as:entropy} is a stochastic equicontinuity condition, which is needed to establish the weak
convergence of our test statistic; see \cite{andrews1994empirical}. Again, this is satisfied by widely
used residual functions such as $m\{y,g(t;\boldsymbol{\theta})\}=y-g(t;\boldsymbol{\theta})$ and $m\{y,g(t;\boldsymbol{\theta})\}=\tau-\mathds{1}\{y<g(t;\boldsymbol{\theta})\}$ discussed in Section \ref{sec:framework}.

To aid presentation of the asymptotic properties of the test statistic, define the following quantities: 
\begin{align*}
\phi(T_i,\boldsymbol{X}_i;t):=&\pi_0(T_i,\boldsymbol{X}_i)\cdot \mathscr{H}(T_i,t)\cdot \mathbb{E}[m\left\{Y_i;g(T_i;{\boldsymbol{\theta}}^*)\right\}|T_i,\boldsymbol{X_i}]\\
&-\mathbb{E}[\pi_0(T_i,\boldsymbol{X}_i)m\left\{Y_i;g(T_i;{\boldsymbol{\theta}}^*)\right\}\cdot \mathscr{H}(T_i,t)|\boldsymbol{X}_i],
\end{align*}
and
\begin{align}
\psi(T_i,\boldsymbol{X}_i,Y_i;t):=&\mathbb{E}\left[{\pi}_0(T_i,\boldsymbol{X}_i)\cdot\frac{\partial}{\partial g}\mathbb{E}[m\left\{Y_i;g(T_i;\boldsymbol{\theta}^*)\right\}|T_i,\boldsymbol{X}_i]\cdot \nabla_{\boldsymbol{\theta}}g(T_i;\boldsymbol{\theta}^*)^\top\mathscr{H}(T_i,t)\right]\nonumber\\
&\times \bigg\{\mathbb{E}\left[{\pi}_0(T_i,\boldsymbol{X}_i)\cdot \frac{\partial}{\partial g}\mathbb{E}[m\left\{Y_i;g(T_i;\boldsymbol{\theta}^*)\right\}|T_i,\boldsymbol{X}_i]\cdot\nabla_{\boldsymbol{\theta}}g(T_i;\boldsymbol{\theta}^*) w(T_i;\boldsymbol{\theta}^*)^\top\right]\notag\nonumber\\
&\qquad \cdot \mathbb{E}\left[{\pi}_0(T_i,\boldsymbol{X}_i)\cdot \frac{\partial}{\partial g}\mathbb{E}[m\left\{Y_i;g(T_i;\boldsymbol{\theta}^*)\right\}|T_i,\boldsymbol{X}_i]\cdot w(T_i;\boldsymbol{\theta}^*)\nabla^\top_{\boldsymbol{\theta}}g(T_i;\boldsymbol{\theta}^*)\right]\bigg\}^{-1}\notag\nonumber\\
&\times\mathbb{E}\left[{\pi}_0(T_i,\boldsymbol{X}_i)\cdot \frac{\partial}{\partial g}\mathbb{E}[m\left\{Y_i;g(T_i;\boldsymbol{\theta}^*)\right\}|T_i,\boldsymbol{X}_i]\cdot\nabla_{\boldsymbol{\theta}}g(T_i;\boldsymbol{\theta}^*)w(T_i;\boldsymbol{\theta}^*)^\top\right]\nonumber\\
&\times\bigg\{\pi_0(T_i,\boldsymbol{X}_i)m\left\{Y_i;g(T_i;\boldsymbol{\theta}^*)\right\}w(T_i;\boldsymbol{\theta}^*)\notag\\
& \qquad-{\pi}_0(T_i,\boldsymbol{X}_i)w(T_i;{\boldsymbol{\theta}}^*)\cdot\mathbb{E}[m\left\{Y_i;g(T_i;\boldsymbol{\theta}^*)\right\}|T_i,\boldsymbol{X}_i]\notag\nonumber\\
&\qquad+\mathbb{E}[{\pi}_0(T_i,\boldsymbol{X}_i)w(T_i;{\boldsymbol{\theta}}^*)m\left\{Y_i;g(T_i;\boldsymbol{\theta}^*)\right\}|\boldsymbol{X}_i]\bigg\},\label{psidef}
\end{align}
and
\begin{align}\label{def:eta}
\eta(T_i,\boldsymbol{X}_i,Y_i;t):=U_i\mathscr{H}(T_i,t)-\phi(T_i,\boldsymbol{X}_i;t)-\psi(T_i,\boldsymbol{X}_i,Y_i;t).
\end{align}
The next theorem establishes the weak convergence of $\widehat{J}_N(\cdot)$ and $\widehat{CM}_N$ under $H_0$.
\begin{theorem}\label{Asymptotic:H0}
	Suppose that  Assumptions \ref{as:TYindep}-\ref{as:entropy} and Assumptions~A.1-A.4 listed in section~A of the supplementary file hold; then, under $H_0$,
	\begin{align*}
	&(i)\quad \widehat{J}_N(t)=\frac{1}{\sqrt{N}}\sum_{i=1}^N\eta(T_i,\boldsymbol{X}_i,Y_i;t) +o_P(1)\ \text{holds uniformly over}\ t\in\mathcal{T},\\
	&(ii)\quad \widehat{J}_N(\cdot) \ \text{converges weakly to} \ J_{\infty}(\cdot) \ \text{in} \  L_2\{\mathcal{T},dF_T(t)\}\,,\nonumber
	\end{align*}
	where $J_{\infty}$ is a Gaussian process with zero mean and covariance function given by
	\begin{align*}
	\Sigma(t,t')=\mathbb{E}\left\{\eta(T_i,\boldsymbol{X}_i,Y_i;t)\eta(T_i,\boldsymbol{X}_i,Y_i;t')\right\}.
	\end{align*}
	Furthermore,
	\begin{align*}
	&(iii) \quad \widehat{CM}_N \ \text{converges to} \ \int \{J_{\infty}(t)\}^2dF_T(t) \ \text{in distribution},\\
	&(iv) \quad \widehat{KS}_N \ \text{converges to} \ \sup_{t\in\mathcal{T}}|J_{\infty}(t)| \ \text{in distribution}.
	\end{align*}
\end{theorem}
The proof of Theorem \ref{Asymptotic:H0} is relegated to section~B in the supplementary file. Similar to \cite{bierens1997asymptotic}, \cite{chen1999consistent}, it can be shown that $\int \{J_{\infty}(t)\}^2dF_T(t)$ can be written as an infinite sum of weighted (independent) $\chi_1^2$ random variables with weights depending on the unknown distribution of $(T_i,\boldsymbol{X}_i,Y_i)$. Hence, it is difficult to obtain the exact critical values. We suggest a simulation method to approximate the critical values for the null limiting distribution of $\widehat{CM}_N$; see Section~\ref{sec:approximation}.

The effect of the vector $w(T;\boldsymbol{\theta})$ on the asymptotic property of our test statistic is reflected in the term $\psi(T_i,\boldsymbol{X}_i,Y_i,t)$. It is unclear which choice of $w(T;\boldsymbol{\theta})$ would minimize the variance of $\Sigma(t,t)$. A common choice is  $w(T;\boldsymbol{\theta}) = \nabla_{\boldsymbol{\theta}}g(T;\boldsymbol{\theta})$. Then, the second and fourth terms of $\psi(T_i,\boldsymbol{X}_i,Y_i,t)$ in \eqref{psidef} are canceled out, which also simplifies the calculation approximating the null limiting distribution in practice.

The next theorem shows that the proposed test statistic is
more efficient than the infeasible test statistic constructed by using the true $\pi_0(T,\boldsymbol{X})$. Suppose that $\pi_0(T,\boldsymbol{X})$ was known, let $\widehat{\boldsymbol{\theta}}_0$ be the estimator of $\boldsymbol{\theta}^*$ constructed by using the true ratio function $\pi_0(T,\boldsymbol{X})$, which is defined by minimizing the following criterion function:
\begin{align*}
\widehat{\boldsymbol{\theta}}_0=\arg\min_{\boldsymbol{\theta}\in\Theta}\|M_N(\boldsymbol{\theta},\pi_0)\|.
\end{align*}
The infeasible test statistic for $H_0$ is then based on
\begin{align*}
\widehat{J}_{0}(t)=\frac{1}{\sqrt{N}}\sum_{i=1}^N	\widehat{U}_{0i}\mathscr{H}(T_i,t), \ \text{where}\ \widehat{U}_{0i}=\pi_{0}(T_i,\boldsymbol{X}_i)m\left\{Y_i;g(T_i;\widehat{\boldsymbol{\theta}}_0)\right\}.
\end{align*}
Let
\begin{align*}
\psi_0(T_i,\boldsymbol{X}_i,Y_i;t):=&\mathbb{E}\left[{\pi}_0(T_i,\boldsymbol{X}_i)\cdot\frac{\partial}{\partial g}\mathbb{E}[m\left\{Y_i;g(T_i;\boldsymbol{\theta}^*)\right\}|T_i,\boldsymbol{X}_i]\cdot \nabla_{\boldsymbol{\theta}}g(T_i;\boldsymbol{\theta}^*)^\top\mathscr{H}(T_i,t)\right]\\
&\times \bigg\{\mathbb{E}\left[{\pi}_0(T_i,\boldsymbol{X}_i)\cdot \frac{\partial}{\partial g}\mathbb{E}[m\left\{Y_i;g(T_i;\boldsymbol{\theta}^*)\right\}|T_i,\boldsymbol{X}_i]\cdot\nabla_{\boldsymbol{\theta}}g(T_i;\boldsymbol{\theta}^*) w(T_i;\boldsymbol{\theta}^*)^\top\right]\notag\\
&\qquad \cdot \mathbb{E}\left[{\pi}_0(T_i,\boldsymbol{X}_i)\cdot \frac{\partial}{\partial g}\mathbb{E}[m\left\{Y_i;g(T_i;\boldsymbol{\theta}^*)\right\}|T_i,\boldsymbol{X}_i]\cdot w(T_i;\boldsymbol{\theta}^*)\nabla^\top_{\boldsymbol{\theta}}g(T_i;\boldsymbol{\theta}^*)\right]\bigg\}^{-1}\notag\\
&\times\mathbb{E}\left[{\pi}_0(T_i,\boldsymbol{X}_i)\cdot \frac{\partial}{\partial g}\mathbb{E}[m\left\{Y_i;g(T_i;\boldsymbol{\theta}^*)\right\}|T_i,\boldsymbol{X}_i]\cdot\nabla_{\boldsymbol{\theta}}g(T_i;\boldsymbol{\theta}^*)w(T_i;\boldsymbol{\theta}^*)^\top\right]\\
&\times\pi_0(T_i,\boldsymbol{X}_i)m\left\{Y_i;g(T_i;\boldsymbol{\theta}^*)\right\}w(T_i;\boldsymbol{\theta}^*)\notag,
\end{align*} 
and
\begin{align*}
\eta_0(T_i,\boldsymbol{X}_i,Y_i;t):=U_i\mathscr{H}(T_i,t)-\psi_0(T_i,\boldsymbol{X}_i,Y_i;t).
\end{align*}

The following theorem establishes the weak convergence of $\widehat{J}_0(\cdot)$ under $H_0$ and shows that the asymptotic variance of the proposed test statistic $\widehat{J}_N(t)$ is smaller than that of  $\widehat{J}_0(t)$ for any $t\in \mathcal{T}$. 
\begin{theorem}\label{thm:asymp_pi0}
	Suppose that Assumptions \ref{as:var_Y}-\ref{as:entropy} hold, then under $H_0$,
	\begin{align*}
	&(i)\quad \widehat{J}_0(t)=\frac{1}{\sqrt{N}}\sum_{i=1}^N\eta_0(T_i,\boldsymbol{X}_i,Y_i;t) +o_P(1)\ \text{holds uniformly over}\ t\in\mathcal{T},\\
	&(ii)\quad \widehat{J}_0(\cdot) \ \text{converges weakly to} \ J_{0,\infty}(\cdot) \ \text{in} \  L_2\{\mathcal{T},dF_T(t)\},
	\end{align*}
	where $J_{0,\infty}$ is a Gaussian process with zero mean and covariance function given by
	\begin{align*}
	\Sigma_0(t,t')=\mathbb{E}\left\{\eta_0(T_i,\boldsymbol{X}_i,Y_i;t)\eta_0(T_i,\boldsymbol{X}_i,Y_i;t')\right\}.
	\end{align*}
	Furthermore, $\Sigma_0(t,t)>\Sigma(t,t)$ for any $t\in \mathcal{T}$.
\end{theorem}
The proof of Theorem~\ref{thm:asymp_pi0} is presented in section~C in the supplementary file. In the estimation of the average treatment effects with binary and multiple treatments, it is a well-known paradox that using a nonparametric estimated propensity score is more efficient than using the true one; see \cite{Hirano03}, \cite{chan2016globally}, and \cite{lee2018efficient} among others. Theorem~\ref{thm:asymp_pi0} shows that this is also the case for continuous treatments.
\subsection{Special cases}
This section discusses two important special continuous treatment effect models, the average and quantile continuous treatment models. In the case of testing for the average dose-response model, that is,
\begin{align}\label{null_mean}
H_0: \exists\ \text{some}\ \boldsymbol{\theta}^*\in \Theta\subset \mathbb{R}^p,\ \text{s.t.}\ \mathbb{E}\{Y^*(t)\}=g(t;\boldsymbol{\theta}^*)\ \text{for all $t\in\mathcal{T}$},
\end{align}
against the alternative hypothesis
\begin{align*}
H_1: \nexists\ \text{any}\ \boldsymbol{\theta}\in \Theta\subset \mathbb{R}^p,\ \text{s.t.}\ \mathbb{E}\{Y^*(t)\}=g(t;\boldsymbol{\theta})=0\ \text{for all $t\in\mathcal{T}$}\,,
\end{align*}
$m\left\{Y^*(t);g(t;\boldsymbol{\theta}^*)\right\} = Y^*(t)- g(t;\boldsymbol{\theta}^*)$, $U_i^{ADRF}=\pi_0(T_i,\boldsymbol{X}_i)\{Y_i-g(T_i;\boldsymbol{\theta}^*)\}$ and the test statistics for $H_0$ are
\begin{align*}
\widehat{CM}^{ADRF}_N=\frac{1}{N}\sum_{i=1}^N\{\widehat{J}^{ADRF}_N(T_i)\}^2 \ \text{and} \ \widehat{KS}^{ADRF}_N=\sup_{t\in\mathcal{T}}|\widehat{J}^{ADRF}_N(t)|,
\end{align*}
where
\begin{align*}
\widehat{J}^{ADRF}_N(t)=\frac{1}{\sqrt{N}}\sum_{i=1}^N\widehat{U}^{ADRF}_i\mathscr{H}(T_i,t), \
\widehat{U}^{ADRF}_i=\widehat{\pi}_{K}(T_i,\boldsymbol{X}_i)\left\{Y_i-g(T_i;\widehat{\boldsymbol{\theta }})\right\}.
\end{align*}
In this special case, the notations $\phi(T_i,\boldsymbol{X}_i;t)$, $\psi(T_i,\boldsymbol{X}_i,Y_i;t)$, and $\eta(T_i,\boldsymbol{X}_i,Y_i;t)$ in Theorem \ref{Asymptotic:H0} become
\begin{align*}
\phi^{ADRF}(T_i,\boldsymbol{X}_i;t):=&\pi_0(T_i,\boldsymbol{X}_i)\cdot \mathscr{H}(T_i,t)\cdot \mathbb{E}\{Y_i-g(T_i;{\boldsymbol{\theta}}^*)|T_i,\boldsymbol{X_i}\}\\
&-\mathbb{E}[\pi_0(T_i,\boldsymbol{X}_i)\{Y_i-g(T_i;{\boldsymbol{\theta}}^*)\}\cdot \mathscr{H}(T_i,t)|\boldsymbol{X}_i],
\end{align*}
and
\begin{align*}
\psi^{ADRF}&(T_i,\boldsymbol{X}_i,Y_i;t):=\mathbb{E}\left[{\pi}_0(T_i,\boldsymbol{X}_i)\cdot \nabla_{\boldsymbol{\theta}}g(T_i;\boldsymbol{\theta}^*)^\top\mathscr{H}(T_i,t)\right]\\
&\times \bigg\{ \mathbb{E}\left[{\pi}_0(T_i,\boldsymbol{X}_i) \nabla_{\boldsymbol{\theta}}g(T_i;\boldsymbol{\theta}^*)w(T_i;\boldsymbol{\theta}^*)^\top\right]\cdot\mathbb{E}\left[{\pi}_0(T_i,\boldsymbol{X}_i)w(T_i;\boldsymbol{\theta}^*) \nabla_{\boldsymbol{\theta}}^\top g(T_i;\boldsymbol{\theta}^*)\right]\bigg\}^{-1} \\
&\times\mathbb{E}\left[{\pi}_0(T_i,\boldsymbol{X}_i) \nabla_{\boldsymbol{\theta}}g(T_i;\boldsymbol{\theta}^*)w(T_i;\boldsymbol{\theta}^*)^\top\right]\\
&\times \bigg\{\pi_0(T_i,\boldsymbol{X}_i)w(T_i;\boldsymbol{\theta}^*)Y_i-{\pi}_0(T_i,\boldsymbol{X}_i)w(T_i;{\boldsymbol{\theta}}^*)\cdot\mathbb{E}(Y_i|T_i,\boldsymbol{X}_i)\\
&\qquad+\mathbb{E}[{\pi}_0(T_i,\boldsymbol{X}_i)w(T_i;{\boldsymbol{\theta}}^*)\{Y_i-g(T_i;\boldsymbol{\theta}^*)\}|\boldsymbol{X}_i]\bigg\},
\end{align*} 
and
\begin{align*}
\eta^{ADRF}(T_i,\boldsymbol{X}_i,Y_i;t):=U_i^{ADRF}\mathscr{H}(T_i,t)-\phi^{ADRF}(T_i,\boldsymbol{X}_i;t)-\psi^{ADRF}(T_i,\boldsymbol{X}_i,Y_i;t).
\end{align*}

Then Theorem \ref{Asymptotic:H0} implies the following result. 
\begin{cor}\label{cor:DRF}
	Suppose that Assumptions \ref{as:TYindep}-\ref{as:var_Y} and Assumptions~A.1-A.4 listed in section~A of the supplementary file hold; then, under $H_0$,
	\begin{align*}
	&(i)\quad \widehat{J}^{ADRF}_N(t)=\frac{1}{\sqrt{N}}\sum_{i=1}^N\eta^{ADRF}(T_i,\boldsymbol{X}_i,Y_i;t) +o_P(1)\ \text{holds uniformly over}\ t\in\mathcal{T},\\
	&(ii) \quad \widehat{J}^{ADRF}_N(\cdot) \ \text{converges weakly to} \ J^{ADRF}_{\infty}(\cdot) \ \text{in} \  L_2\{\mathcal{T},dF_T(t)\},
	\end{align*}
	where $J^{ADRF}_{\infty}$ is a Gaussian process with zero mean and covariance function given by
	\begin{align*}
	\Sigma^{ADRF}(t,t')=\mathbb{E}\left\{\eta^{ADRF}(T_i,\boldsymbol{X}_i,Y_i;t)\eta^{ADRF}(T_i,\boldsymbol{X}_i,Y_i;t')\right\}.
	\end{align*}
	Furthermore,
	\begin{align*}
	&(iii) \quad \widehat{CM}^{ADRF}_N \ \text{converges to} \ \int \{J^{ADRF}_{\infty}(t)\}^2dF_T(t) \ \text{in distribution},\\
	&(iv) \quad \widehat{KS}^{ADRF}_N \ \text{converges to} \ \sup_{t\in\mathcal{T}}\left|J^{ADRF}_{\infty}(t)\right| \ \text{in distribution}.
	\end{align*}
\end{cor}

In the case of testing for the quantile dose-response model, that is,
\begin{align}\label{null_quant}
H_0: \exists\ \text{some}\ \boldsymbol{\theta}^*\in \Theta\subset \mathbb{R}^p,\ \text{s.t.}\ F^{-1}_{Y^*(t)}(\tau)=g(t;\boldsymbol{\theta}^*)\ \text{for all $t\in\mathcal{T}$},
\end{align}
against the alternative hypothesis
\begin{align*}
H_1: \nexists\ \text{any}\ \boldsymbol{\theta}\in \Theta\subset \mathbb{R}^p,\ \text{s.t.}\ F^{-1}_{Y^*(t)}(\tau)=g(t;\boldsymbol{\theta})\ \text{for all $t\in\mathcal{T}$}\,,
\end{align*}
$m\left\{Y^*(t);g(t;\boldsymbol{\theta}^*)\right\} = \tau-\mathbbm{1}\{Y^*(t)<g(t;\boldsymbol{\theta}^*)\}$, $U_i^{QDRF}=\pi_0(T_i,\boldsymbol{X}_i)\big[\tau-\mathbbm{1}\{Y_i<g(T_i;\boldsymbol{\theta}^*)\}\big]$, and the test statistics for $H_0$ are
\begin{align*}
\widehat{CM}^{QDRF}_N=\frac{1}{N}\sum_{i=1}^N[\widehat{J}^{QDRF}_N(T_i)]^2 \ \text{and} \ \widehat{KS}^{QDRF}_N=\sup_{t\in\mathcal{T}}\left|\widehat{J}^{QDRF}_N(t)\right|,
\end{align*}
where
\begin{align*}
\widehat{J}^{QDRF}_N(t)=\frac{1}{\sqrt{N}}\sum_{i=1}^N\widehat{U}^{QDRF}_i\mathscr{H}(T_i,t), \
\widehat{U}^{QDRF}_i=\widehat{\pi}_{K}(T_i,\boldsymbol{X}_i)\left[\tau-\mathbbm{1}\{Y_i<g(t;\widehat{\boldsymbol{\theta}})\}\right].
\end{align*}
Again, in this special case, the notations $\phi(T_i,\boldsymbol{X}_i;t)$, $\psi(T_i,\boldsymbol{X}_i,Y_i;t)$, and $\eta(T_i,\boldsymbol{X}_i,Y_i;t)$ in Theorem \ref{Asymptotic:H0} become
\begin{align*}
\phi^{QDRF}(T_i,\boldsymbol{X}_i;t):=&\pi_0(T_i,\boldsymbol{X}_i)\cdot \mathbb{E}\Big(\left[\tau-\mathbbm{1}\{Y_i<g(T_i;\boldsymbol{\theta}^*)\}\right]\cdot \mathscr{H}(T_i,t)|T_i,\boldsymbol{X_i}\Big)\\
&-\mathbb{E}\Big\{\pi_0(T_i,\boldsymbol{X}_i)\left[\tau-\mathbbm{1}\{Y_i<g(T_i;\boldsymbol{\theta}^*)\}\right]\cdot \mathscr{H}(T_i,t)|\boldsymbol{X}_i\Big\},
\end{align*}
and
\begin{align*}
\psi^{QDRF}(T_i,\boldsymbol{X}_i,Y_i;t):=&\mathbb{E}\left[{\pi}_0(T_i,\boldsymbol{X}_i)\cdot f_{Y|T,X}\{g(T_i;\boldsymbol{\theta}^*)|T_i,\boldsymbol{X}_i\} \cdot \nabla_{\boldsymbol{\theta}}g(T_i;\boldsymbol{\theta}^*)^\top\mathscr{H}(T_i,t)\right]\\
&\times \bigg\{\mathbb{E}\left[{\pi}_0(T_i,\boldsymbol{X}_i)\cdot f_{Y|T,X}\{g(T_i;\boldsymbol{\theta}^*)|T_i,\boldsymbol{X}_i\} \cdot\nabla_{\boldsymbol{\theta}}g(T_i;\boldsymbol{\theta}^*)w(T_i;\boldsymbol{\theta}^*)^\top\right]\\
&\quad \cdot\mathbb{E}\left[{\pi}_0(T_i,\boldsymbol{X}_i)\cdot f_{Y|T,X}\{g(T_i;\boldsymbol{\theta}^*)|T_i,\boldsymbol{X}_i\} \cdot w(T_i;\boldsymbol{\theta}^*)\nabla_{\boldsymbol{\theta}}g(T_i;\boldsymbol{\theta}^*)^\top\right]\bigg\}^{-1} \\
&\times\mathbb{E}\left[{\pi}_0(T_i,\boldsymbol{X}_i)\cdot f_{Y|T,X}\{g(T_i;\boldsymbol{\theta}^*)|T_i,\boldsymbol{X}_i\} \cdot\nabla_{\boldsymbol{\theta}}g(T_i;\boldsymbol{\theta}^*)w(T_i;\boldsymbol{\theta}^*)^\top\right]\\
&\times \bigg\{-\pi_0(T_i,\boldsymbol{X}_i)w(T_i;\boldsymbol{\theta}^*)\mathbbm{1}\{Y_i<g(T_i;\boldsymbol{\theta}^*)\}\\
&\qquad +{\pi}_0(T_i,\boldsymbol{X}_i)w(T_i;{\boldsymbol{\theta}}^*)\cdot\mathbb{E}[\mathbbm{1}\{Y_i<g(T_i;\boldsymbol{\theta}^*)\}|T_i,\boldsymbol{X}_i]\\
&\qquad+\mathbb{E}\Big[{\pi}_0(T_i,\boldsymbol{X}_i)w(T_i;{\boldsymbol{\theta}}^*)\left[\tau-\mathbbm{1}\{Y_i<g(T_i;\boldsymbol{\theta}^*)\}\right]|\boldsymbol{X}_i\Big]\bigg\}\,,
\end{align*} 
and
\begin{align*}
\eta^{QDRF}(T_i,\boldsymbol{X}_i,Y_i;t):=U_i^{QDRF}\mathscr{H}(T_i,t)-\phi^{QDRF}(T_i,\boldsymbol{X}_i;t)-\psi^{QDRF}(T_i,\boldsymbol{X}_i,Y_i;t).
\end{align*}
Then, Theorem \ref{Asymptotic:H0} implies the following result. 
\begin{cor}\label{cor:QDRF}
	Suppose that Assumptions \ref{as:TYindep}-\ref{as:var_Y} and Assumptions~A.1-A.4 listed in section~A of the supplementary file hold; then, under $H_0$,
	\begin{align*}
	&(i)\quad \widehat{J}^{QDRF}_N(t)=\frac{1}{\sqrt{N}}\sum_{i=1}^N\eta^{QDRF}(T_i,\boldsymbol{X}_i,Y_i;t) +o_P(1)\ \text{holds uniformly over}\ t\in\mathcal{T},\\
	&(ii) \quad \widehat{J}^{QDRF}_N(\cdot) \ \text{converges weakly to} \ J^{QDRF}_{\infty}(\cdot) \ \text{in} \  L_2\{\mathcal{T},dF_T(t)\},
	\end{align*}
	where $J^{QDRF}_{\infty}$ is a Gaussian process with zero mean and covariance function given by
	\begin{align*}
	\Sigma^{QDRF}(t,t')=\mathbb{E}\left\{\eta^{QDRF}(T_i,\boldsymbol{X}_i,Y_i;t)\eta^{QDRF}(T_i,\boldsymbol{X}_i,Y_i;t')\right\}.
	\end{align*}
	Furthermore,
	\begin{align*}
	&(iii) \quad \widehat{CM}^{QDRF}_N \ \text{converges to} \ \int \{J^{QDRF}_{\infty}(t)\}^2dF_T(t) \ \text{in distribution},\\
	&(iv) \quad \widehat{KS}^{QDRF}_N \ \text{converges to} \ \sup_{t\in\mathcal{T}} \left|J^{QDRF}_{\infty}(t)\right| \ \text{in distribution}.	
	\end{align*}
\end{cor}

\subsection{Asymptotic properties under the fixed and local alternative hypothesis}\label{sec:AsymptoticAlternatives}
This section studies the asymptotic distribution of $\widehat{J}_N(\cdot)$  under the fixed and Pitman local
alternatives. The Pitman local alternative is given by
\begin{align*}
H_L: \mathbb{E}\left[m\left\{Y^*(t);g(t;\boldsymbol{\theta}_N^*)+\frac{1}{\sqrt{N}}\cdot \delta(t)\right\}\right]=0 \ \ \text{for some} \ \boldsymbol{\theta}_N^*\in \Theta \ \ \text{and all} \ t\in\mathcal{T} \,,
\end{align*}
where $\int \{\delta (t)\}^2dF_T(t)<\infty$. With Assumption \ref{as:TYindep}, $H_L$ can be represented by
\begin{align*}
H_L: \mathbb{E}\left[\pi_0(T,\boldsymbol{X})m\left\{Y;g(T;\boldsymbol{\theta}_N^*)+\frac{1}{\sqrt{N}}\cdot \delta(T)\right\}\bigg|T=t\right]=0 \ \ \text{for some} \ \boldsymbol{\theta}_N^*\in \Theta \ \ \text{and all} \ t\in\mathcal{T}\,,
\end{align*}
which deviates from the null model at the rate of $O(N^{-1/2})$.
Let $\boldsymbol{\theta}^*$ be the limit of $\boldsymbol{\theta}^*_N$ as $N\rightarrow\infty$, hence it solves the following equation:
\begin{align*}
\mathbb{E}\left[\pi_0(T,\boldsymbol{X})m\left\{Y;g(T;\boldsymbol{\theta}^*)\right\}\bigg|T=t\right]=0\ \ \text{for all} \ t\in\mathcal{T}.
\end{align*}
Define
\begin{align*}
\mu(t):=& \mathbb{E}\left[{\pi}_0(T_i,\boldsymbol{X}_i)\cdot\frac{\partial}{\partial g}\mathbb{E}[m\left\{Y_i;g(T_i;\boldsymbol{\theta}^*)\right\}|T_i,\boldsymbol{X}_i]\cdot \nabla_{\boldsymbol{\theta}}g(T_i;\boldsymbol{\theta}^*)^\top\mathscr{H}(T_i,t)\right]\\
&\times\Bigg\{\mathbb{E}\left[\pi_0(T_i,\boldsymbol{X}_i)\cdot \frac{\partial}{\partial g}\mathbb{E}\left[m\left\{Y_i;g(T_i;{\boldsymbol{\theta}}^*)\right\}|T_i,\boldsymbol{X}_i\right]\cdot \nabla_{\boldsymbol{\theta}} g(T_i;\boldsymbol{\theta}^*)w(T_i;\boldsymbol{\theta}^*)^\top \right]\\
&\quad\quad\cdot\mathbb{E}\left[\pi_0(T_i,\boldsymbol{X}_i)\cdot \frac{\partial}{\partial g}\mathbb{E}\left[m\left\{Y_i;g(T_i;{\boldsymbol{\theta}}^*)\right\}|T_i,\boldsymbol{X}_i\right]\cdot w(T_i;\boldsymbol{\theta}^*) \nabla_{\boldsymbol{\theta}} g(T_i;\boldsymbol{\theta}^*)^\top \right]\Bigg\}^{-1} \\
&\times\mathbb{E}\left[\pi_0(T_i,\boldsymbol{X}_i)\cdot \frac{\partial}{\partial g}\mathbb{E}\left[m\left\{Y_i;g(T_i;{\boldsymbol{\theta}}^*)\right\}|T_i,\boldsymbol{X}_i\right]\cdot \nabla_{\boldsymbol{\theta}} g(T_i;\boldsymbol{\theta}^*)w(T_i;\boldsymbol{\theta}^*)^\top \right]\\
&\times \mathbb{E}\left[\pi_0(T_i,\boldsymbol{X}_i)\cdot \frac{\partial}{\partial g}\mathbb{E}\left[m\left\{Y_i;g(T_i;{\boldsymbol{\theta}}_N^*)\right)\}\big|T_i,\boldsymbol{X}_i\right]\cdot \delta(T_i)\cdot w(T_i;{\boldsymbol{\theta}^*})\right].
\end{align*}

The following theorem gives the asymptotic distribution of $\widehat{J}_N(\cdot)$ under the local alternative $H_L$ and the fixed alternative $H_1$. 
\begin{theorem}\label{Asymptotic:H_L}
	Suppose that Assumptions \ref{as:TYindep}-\ref{as:entropy} and Assumptions~A.1-A.4 listed in section~A of the supplementary file hold. Under the local alternative hypothesis $H_L$,
	\begin{align}
	&(i)\quad \widehat{J}_N(t)=\frac{1}{\sqrt{N}}\sum_{i=1}^N\eta(T_i,\boldsymbol{X}_i,Y_i;t) + \mu(t)+o_P(1)\ \text{holds uniformly over}\ t\in\mathcal{T},\label{JNasymptotics_L}\\
	&(ii)\quad \widehat{J}_N(\cdot) \ \text{converges weakly to} \ J_{\infty,\mu}(\cdot) \ \text{in} \  L_2\{\mathcal{T},dF_T(t)\},\nonumber
	\end{align}
	where $J_{\infty,\mu}$ is a Gaussian process with mean function $\mu(t)$ and covariance function given by
	\begin{align*}
	\Sigma(t,t')=\mathbb{E}\left\{\eta(T_i,\boldsymbol{X}_i,Y_i;t)\eta(T_i,\boldsymbol{X}_i,Y_i;t')\right\}.
	\end{align*}
	Under the fixed $H_1$,
	\begin{align*}
	&(iii)\quad \frac{1}{\sqrt{N}}\widehat{J}_N(\cdot) \ \text{converges to} \ \mu_1(\cdot ) \ \text{in probability in} \ L^2(\mathcal{T},dt),
	\end{align*}
	where $\mu_1(t):=\mathbb{E}\left[\pi_0(T_i,\boldsymbol{X}_i)m\left\{Y_i;g(T_i;{\boldsymbol{\theta}}^*)\right\}\mathscr{H}(T_i,t)\right]$.
\end{theorem}
Comparing Theorem~\ref{Asymptotic:H_L}~(ii) to Theorem~\ref{Asymptotic:H0}~(ii), we see that our test statistic is able to detect the local alternatives deviated from the null model at the rate of $O(N^{-1/2})$.

\section{Approximation for the null limiting distribution}\label{sec:approximation}
We know from Theorem \ref{Asymptotic:H0} that
$\widehat{CM}_{N}$ converges in distribution to $\int\{J_{\infty}%
(t)\}^{2}\,dF_{T}(t)$. Using techniques similar to those in
\cite{bierens1997asymptotic} and \cite{chen1999consistent}, one can show that
$\int\{J_{\infty}(t)\}^{2}\,dF_{T}(t)$ is an infinite sum of weighted
(independent) $\chi_{1}^{2}$ random variables, where the weights depend on the
unknown distribution of the $(\boldsymbol{X}_{i},T_{i},Y_{i})$'s (see also
\citealp{li2003consistent}). Obtaining the exact critical values is difficult
and we here propose a simulation method to approximate the null limiting
distribution. The method is a special case of the \emph{exchangeable
	bootstrap}
\citep{praestgaard1993exchangeably,van1996weak,chernozhukov2013inference,Donald2014Estimation}.
Specifically, we first generate $B$ sets of $N$ independent standard normal
random variables $w_{1,b},\ldots,w_{N,b}$, for $b=1,\ldots,B$ and $B$ a large
enough integer. Then we define
\begin{align}
\widehat{J}_{N,b}^{*}(t) = \frac{1}{\sqrt{N}} \sum^{N}_{i=1}w_{i,b}%
\widehat{\eta}(T_{i},\boldsymbol{X}_{i},Y_{i};t)\,,
\end{align}
where $\widehat{\eta}(T_{i},\boldsymbol{X}_{i},Y_{i};t) = \widehat{U}%
_{i}\mathscr{H}(T_{i},t) - \widehat{\phi}(T_{i},\boldsymbol{X}_{i};t) -
\widehat{\psi}(T_{i},\boldsymbol{X}_{i},Y_{i};t)$, with $\widehat{\phi}%
(T_{i},\boldsymbol{X}_{i};t)$ and $\widehat{\psi}(T_{i},\boldsymbol{X}%
_{i},Y_{i};t)$ respectively some consistent nonparametric plug-in estimators
of $\phi(T_{i},\boldsymbol{X}_{i};t)$ and $\psi(T_{i},\boldsymbol{X}_{i}%
,Y_{i};t)$ defined above in Theorem \ref{Asymptotic:H0}, for example the
additive penalized spline estimator(see \citealp{ruppert2003} for example) or the series estimator used in \cite{Donald2014Estimation}. 

It is easy to see that $\mathbb{E}^{*}\{w_{i,b}\widehat{\eta}(T_{i}%
,\boldsymbol{X}_{i},Y_{i};t)\} = 0$ and $\mathbb{E}^{*}\{w^{2}_{i,b}%
\widehat{\eta}(T_{i},\boldsymbol{X}_{i},Y_{i};t)\widehat{\eta}(T_{i}%
,\boldsymbol{X}_{i},Y_{i};$ $t^{\prime})\}= \widehat{\eta}(T_{i}%
,\boldsymbol{X}_{i},Y_{i};t)\widehat{\eta}(T_{i},\boldsymbol{X}_{i}%
,Y_{i};t^{\prime})$, for $i=1,\ldots,N$, $b=1,\ldots,B$ and all $t,t^{\prime
}\in\mathcal{T}$, where $\mathbb{E}^{*}\{\cdot\}$ is the conditional
expectation given the data $(T_{i},\boldsymbol{X}_{i},Y_{i})_{i=1}^{N}$.
Because $\widehat{\eta}$ is a consistent estimator of $\eta$, then
$\widehat{J}^{*}_{N,b}(\cdot)$ has the same asymptotic behavior as $\widehat
{J}_{N}(\cdot)$ for $b=1,\ldots,B$. Then, we can approximate the limiting
distributions of $\widehat{CM}_{N}$ and  $\widehat{KS}_{N}$ under $H_{0}$, respectively, by
\[
\widehat{CM}_{N,b}^{*} = \frac{1}{N}\sum^{N}_{i=1}\big\{\widehat{J}_{N,b}%
^{*}(T_{i})\big\}^{2} 
\quad \text{and} \quad 
\widehat{KS}_{N,b}^{*} = \sup_{t\in\mathcal{T}}\big|\widehat{J}_{N,b}%
^{*}(t)\big|,
\]
for $b=1,\ldots,B$. That is, we can approximate the $p$-value for the CM-type statistic by $B^{-1}%
\sum^{B}_{b=1}\mathbbm{1}(\widehat{CM}_{N,b}^{*}\geq\widehat{CM}_{N})$ and that for the KS-type statistic by $B^{-1}%
\sum^{B}_{b=1}\mathbbm{1}(\widehat{KS}_{N,b}^{*}\geq\widehat{KS}_{N})$.

\section{Numerical studies}\label{sec:numerical}
\subsection{Choosing $K_1$ and $K_2$}\label{CV}
The large-sample properties of the proposed estimator
hold for a range of values of $K_1$ and $K_2$. This presents a dilemma for
applied researchers, who have only one finite sample. Too little smoothing
yields a large variance and too much smoothing yields a large bias.
Therefore, applied researchers would benefit from guidance on the
choice of $K_1$ and $K_2$. In this section, we propose a cross-validation
method for choosing the smoothing parameters $K_1$ and $K_2$. Specifically, we split the data set into $F$ sets (say $F=5$ or 10), and select $K_1$ and $K_2$ that minimize the following quantity
\begin{equation}
CV(K_1,K_2) = \sum^{F}_{j=1} \left[\frac{1}{|S_j|}\sum_{k\in S_j} \widehat{\pi}^{(-j)}_K(T_k,\boldsymbol{X}_k)m\left\{Y_k; g\left(T_k;\widehat{\boldsymbol{\theta }}^{(-j)}\right)\right\}\right]^2,\label{criteria:CV}
\end{equation}
where $S_j$ denotes the $j$th set of data of $T,\boldsymbol{X}$ and $Y$, $|S_j|$ denotes the number of individuals in the set $S_j$, and for $j=1,\ldots,F$,
$$
\widehat{\boldsymbol{\theta }}^{(-j)} = \arg \min_{\boldsymbol{\theta}\in\Theta}\left\|M_N^{(-j)}\left(\boldsymbol{\theta},\widehat{\pi}_{K}^{(-j)}\right)\right\|\,,
$$
where 
$$
M_N^{(-j)}\left(\boldsymbol{\theta},\widehat{\pi}_{K}^{(-j)}\right):= \frac{1}{N}\sum_{i\notin S_j}\widehat{\pi}_{K}^{(-j)}(T_{i},\boldsymbol{X}_{i})m\{Y_i;g(T_i,\boldsymbol{\theta})\}w(T_i;\boldsymbol{\theta})\,,
$$
with $\widehat{\pi}_{K}^{(-j)}(T_{i},\boldsymbol{X}_{i})$ obtained in a method identical to that introduced in Section~\ref{sec:framework} via \eqref{def:pihat_dual} and \eqref{def:G^hat}, but excluding samples in $S_j$.
\subsection{Simulation study}\label{sec:simulation}
To assess the performance of our specification test method, we conducted Monte Carlo simulation studies on the following four data generating processes (DGPs):
\begin{eqnarray*}
	&&\textbf{DGP0-L} \quad T=1+0.2X+\xi, \quad \text{and} \quad Y = 1+X+T+\epsilon,\\
	&&\textbf{DGP0-NL} \quad T=0.1X^2+\xi, \quad \text{and} \quad Y = X^2+T+\epsilon,\\
	&&\textbf{DGP1-L} \quad T=1+0.2X+\xi, \quad \text{and} \quad Y = 1+X+0.1T^3+\epsilon,\\
	&&\textbf{DGP1-NL} \quad T=0.1X^2+\xi, \quad \text{and} \quad Y = X^2+0.2T^3+\epsilon,
\end{eqnarray*}
where $\xi$ and $\epsilon$ are independent standard normal random variables, and $X$ is a uniform random variable supported on $[0,1]$. For all the four scenarios, we considered the two-sided hypothesis testing in \eqref{null_1}, where $m\{Y^*(t);g(t;\boldsymbol{\theta}^*)\}=Y^*(t)-g(t;\boldsymbol{\theta}^*)$ (average) and  $m\{Y^*(t);g(t;\boldsymbol{\theta}^*)\}=0.5 - \mathbbm{1}\{Y^*(t)<g(t;\boldsymbol{\theta}^*)\}$ (median), and
$$
g\{t;(\theta^*_0,\theta^*_1)\} = \theta^*_0 + \theta^*_1 t\,.
$$
We take the vector $w(T;\boldsymbol{\theta})=\nabla_{\boldsymbol{\theta}}g(T;\boldsymbol{\theta})$ and adopt the algorithm of \cite{de2019smoothed} for estimating the quantile dose-response function to overcome the computational difficulties with the discontinuity of the indicator function.

Clearly, $H_0$ is true for {\bf DGP0-L} and {\bf DGP0-NL}, but fails for {\bf DGP1-L} and {\bf DGP1-NL}. For each case, we generated 1000 samples of size 100, 200, and 500. The number of samples for the simulation-based approximation of the limiting process is $B=500$ and the number of folds in the cross-validation \eqref{criteria:CV} was taken to be $F=10$. 
We compared the three commonly used weight functions $\mathscr{H}$ that are mentioned in Section~\ref{sec:framework}, namely logistic, cosine-sine, and indicator functions. Specifically, for the logistic weight function, we took the constant $c=5$.  We tested all models using both CM-type and KS-type statistics. The results of the two methods are similar; here, we present those of the CM-type statistic. Results of the KS-type one can be found in the supplementary materials of this paper.

\begin{table}[t]
	\footnotesize
	\centering
	\caption{Estimated sizes}\label{table:EstimatedSizes}
	\resizebox{\textwidth}{!}{  
		\begin{tabular}{ccc|lll|lll|lll}
			\hline
			\multicolumn{3}{c}{} &  \multicolumn{3}{c}{\text{Logistic}}& \multicolumn{3}{c}{\text{Cosine-Sine}} & \multicolumn{3}{c}{\text{Indicator}} \\
			$m(\cdot)$&\text{Model}&$N$  &1\% & 5\% & 10\% & 1\%& 5\% & 10\% & 1\%& 5\% & 10\%\\
			\hline
			\multirow{6}{*}{Average}&\multirow{3}{*}{\textbf{DGP0-L}}
			&100& 0.020& 0.076& 0.131&0.016&0.071&0.138&0.014&0.055&0.115\\
			&&200& 0.015& 0.050& 0.105&0.016&0.056&0.118&0.011&0.057&0.114\\
			&&500& 0.011& 0.051& 0.109&0.012&0.050&0.101&0.014&0.049&0.111\\
			\cline{2-12}
			&\multirow{3}{*}{\textbf{DGP0-NL}}
			&100& 0.033& 0.094& 0.172&0.016&0.080&0.134&0.018&0.061&0.124\\
			&&200& 0.024& 0.079& 0.138&0.010&0.060&0.117&0.010&0.062&0.120\\
			&&500& 0.024& 0.067& 0.129&0.011&0.056&0.101&0.011&0.052&0.118\\
			\hline
			\multirow{6}{*}{Median}&\multirow{3}{*}{\textbf{DGP0-L}}
			&100& 0.036& 0.117& 0.171&0.028&0.090&0.141&0.035&0.100&0.178\\
			&&200& 0.016& 0.076& 0.144&0.017&0.066&0.124&0.018&0.068&0.135\\
			&&500& 0.010& 0.047& 0.104&0.009&0.063&0.117&0.012&0.063&0.120\\
			\cline{2-12}
			&\multirow{3}{*}{\textbf{DGP0-NL}}
			&100& 0.044& 0.135& 0.217&0.014&0.072&0.139&0.022&0.083&0.168\\
			&&200& 0.026& 0.101& 0.160&0.015&0.060&0.113&0.016&0.066&0.141\\
			&&500& 0.020& 0.078& 0.130&0.010&0.054&0.111&0.013&0.061&0.119\\
			\hline
		\end{tabular}}
	\end{table}
	
	Tables~\ref{table:EstimatedSizes} and \ref{table:EstimatedPowers} summarize the empirical rejection probabilities computed at significance levels $1\%, 5\%$, and $10\%$ for each case, which respectively show the estimated sizes (DGP0-L and DGP0NL) and the estimated powers (DGP1-L and DGP1-NL) of our test method.

	\begin{table}[t]
		\footnotesize
		\centering
		\caption{Estimated power}\label{table:EstimatedPowers}
		\resizebox{\textwidth}{!}{  
			{\begin{tabular}{ccc|ccc|ccc|ccc}
					\hline
					\multicolumn{3}{c}{} &  \multicolumn{3}{c}{\text{Logistic}}& \multicolumn{3}{c}{\text{Cosine-Sine}} & \multicolumn{3}{c}{\text{Indicator}} \\
					$m(\cdot)$&\text{Model}&$N$  &1\% & 5\% & 10\% & 1\%& 5\% & 10\% & 1\%& 5\% & 10\%\\
					\hline
					\multirow{4}{*}{Average}&\multirow{2}{*}{\textbf{DGP1-L}}
					&100& 0.829& 0.931& 0.963&0.566&0.797&0.868&0.697&0.848&0.895\\
					&&200& 0.987& 0.998& 0.999&0.918&0.980&0.994&0.969&0.996&0.998\\
					\cline{2-12}
					&\multirow{2}{*}{\textbf{DGP1-NL}}
					&100& 0.543& 0.739& 0.838&0.531&0.743&0.835&0.458&0.669&0.767\\
					&&200& 0.854& 0.945& 0.972& 0.865& 0.950& 0.971& 0.835& 0.925& 0.953\\
					\hline
					\multirow{4}{*}{Median}&\multirow{2}{*}{\textbf{DGP1-L}}
					&100& 0.529& 0.744& 0.832& 0.259& 0.517& 0.656& 0.391& 0.633& 0.743\\
					&&200&  0.870& 0.957& 0.984& 0.580& 0.818& 0.891& 0.760& 0.903& 0.946\\
					\cline{2-12}
					&\multirow{2}{*}{\textbf{DGP1-NL}}
					&100& 0.250& 0.476& 0.619&0.207&0.438&0.590&0.196&0.407&0.542\\
					&&200& 0.545& 0.769& 0.876& 0.511& 0.762& 0.849& 0.471& 0.710& 0.809\\
					\hline
				\end{tabular}}
			}
		\end{table}

		We can see from Table~\ref{table:EstimatedSizes} that the estimated sizes of our method with cosine-sine and indicator weight functions are quite close to the nominal sizes from $N=100$ to $500$ for all cases. The estimated sizes when using the logistic weight function are obviously over-sized when the sample size is small, especially for nonlinear $\boldsymbol{X}$ cases, but they also improve as the sample size increases and are close to the nominal sizes when $N=500$.
		
		From Table~\ref{table:EstimatedPowers}, we observe that all tests become more and more powerful as $N$ or significance level increases and reach a considerably high power level when $N=200$.
		
		Overall, the simulation studies confirmed our asymptotic theorems and showed that, in practice, the cosine-sine and indicator weight functions might perform better than the logistic one for nonlinear $\boldsymbol{X}$ cases.
		
		\subsection{Real data analysis}\label{sec:RealData}
		In this section, we applied our method to examine the model assumption made on the U.S. presidential campaign data in \cite{Ai_Linton_Motegi_Zhang_cts_treat}. The data have been analyzed several times in the treatment effect literature \citep{Urban_Niebler_2014,Fong_Hazlett_Imai_2018}, where the interest was to explore the casual relationship between advertising and campaign contributions. The treatment of interest is the number of political advertisements aired in each zip code from non-competitive states, which ranges from 0 to 22379 across $N=16265$ zip codes.
		
		The data were first analyzed by \cite{Urban_Niebler_2014}, who used a binary model to compare the campaign contributions of the 5230 zip codes that received more than 1000 advertisements with those of the other 11035 zip codes that received less than 1000 advertisements. Their research suggested that advertising in non-competitive states had a significant casual effect on the level of campaign contributions.
		
		By contrast, \cite{Ai_Linton_Motegi_Zhang_cts_treat} considered the treatment variable (number of political advertisements) as continuous and assumed that 
		\begin{equation}
		\mathbb{E}\{Y^*(t)\} = \beta_1 + \beta_2 t + \beta_3 t^2\,,\label{AiModel}
		\end{equation}
		where the observed outcome $Y^*(T) = \log(\text{Contribution}+1)$ and $T = \log(\#\text{ads}+1)$, where \#ads denotes the number of advertisements. The covariates $\boldsymbol{X}$ considered were
		$$
		\boldsymbol{X} =\begin{bmatrix}
		\log(\text{Population})\\
		\%\text{Age over 65}\\
		\log(\text{Median Income})\\
		\%\text{Hispanic}\\
		\%\text{Black}\\
		\log(\text{Population density}+1)\\
		\%\text{College graduates}\\
		\mathbbm{1}(\text{Can commute to a competitive state})
		\end{bmatrix} \,.
		$$
		The definition of each covariate is almost self-explanatory, and one can refer to \cite{Fong_Hazlett_Imai_2018} for
		more details.
		\cite{Ai_Linton_Motegi_Zhang_cts_treat} found that the 95\% confidence intervals for $\beta_2$ and $\beta_3$ were respectively $[-0.025, 0.232]$ and $[-0.025, 0.001]$, indicating that no significant causal link between advertising and campaign contributions was found from the linear model. Similar results were also reported by \cite{Fong_Hazlett_Imai_2018}. The authors then concluded that such opposing results from binary models and continuous linear models suggested a rather complex relationship between advertising and campaign contributions.
		
		\begin{figure}[t]
			\centering
			\includegraphics[width = .35\textwidth]{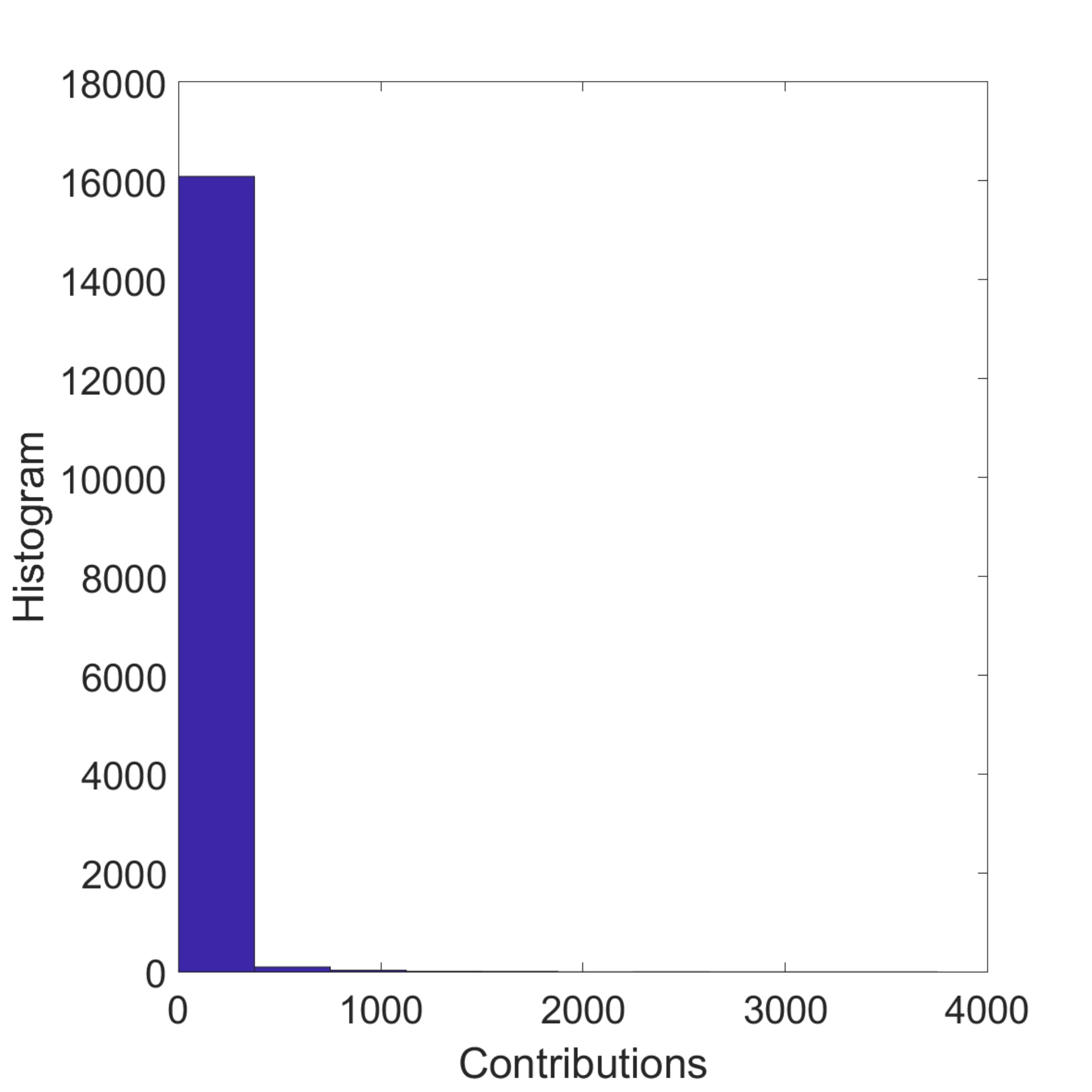}
			\includegraphics[width = .35\textwidth]{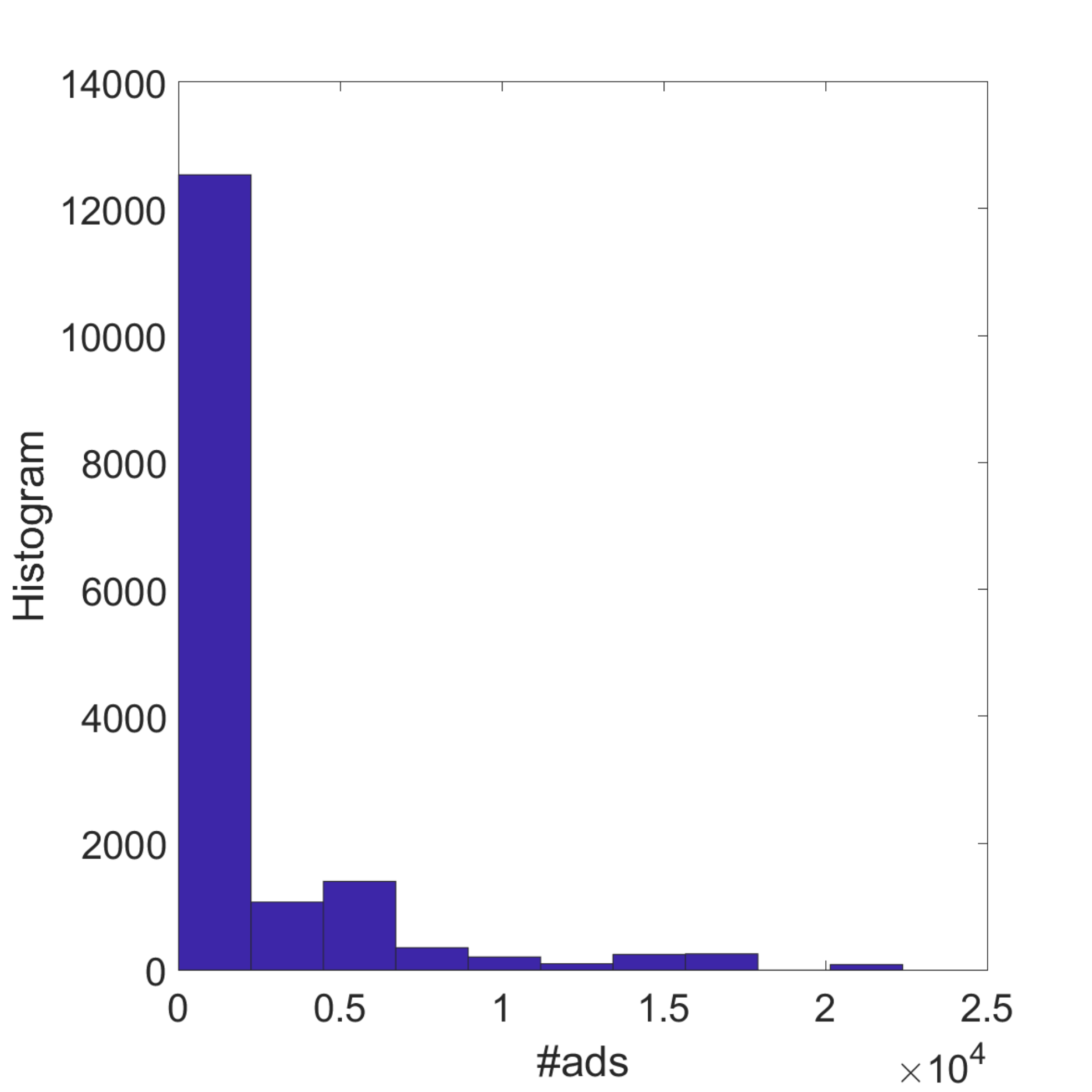}
			
			\includegraphics[width = .35\textwidth]{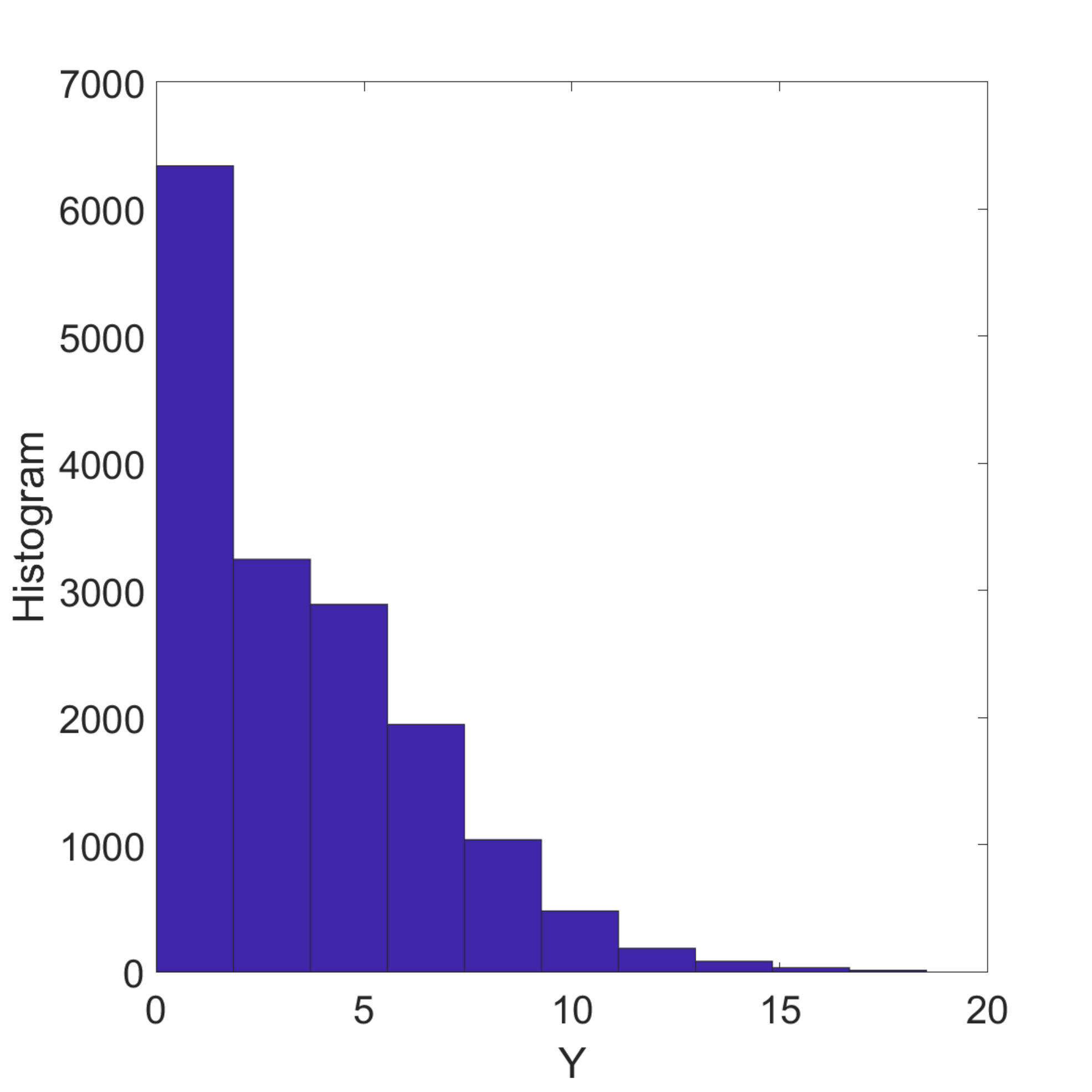}
			\includegraphics[width = .35\textwidth]{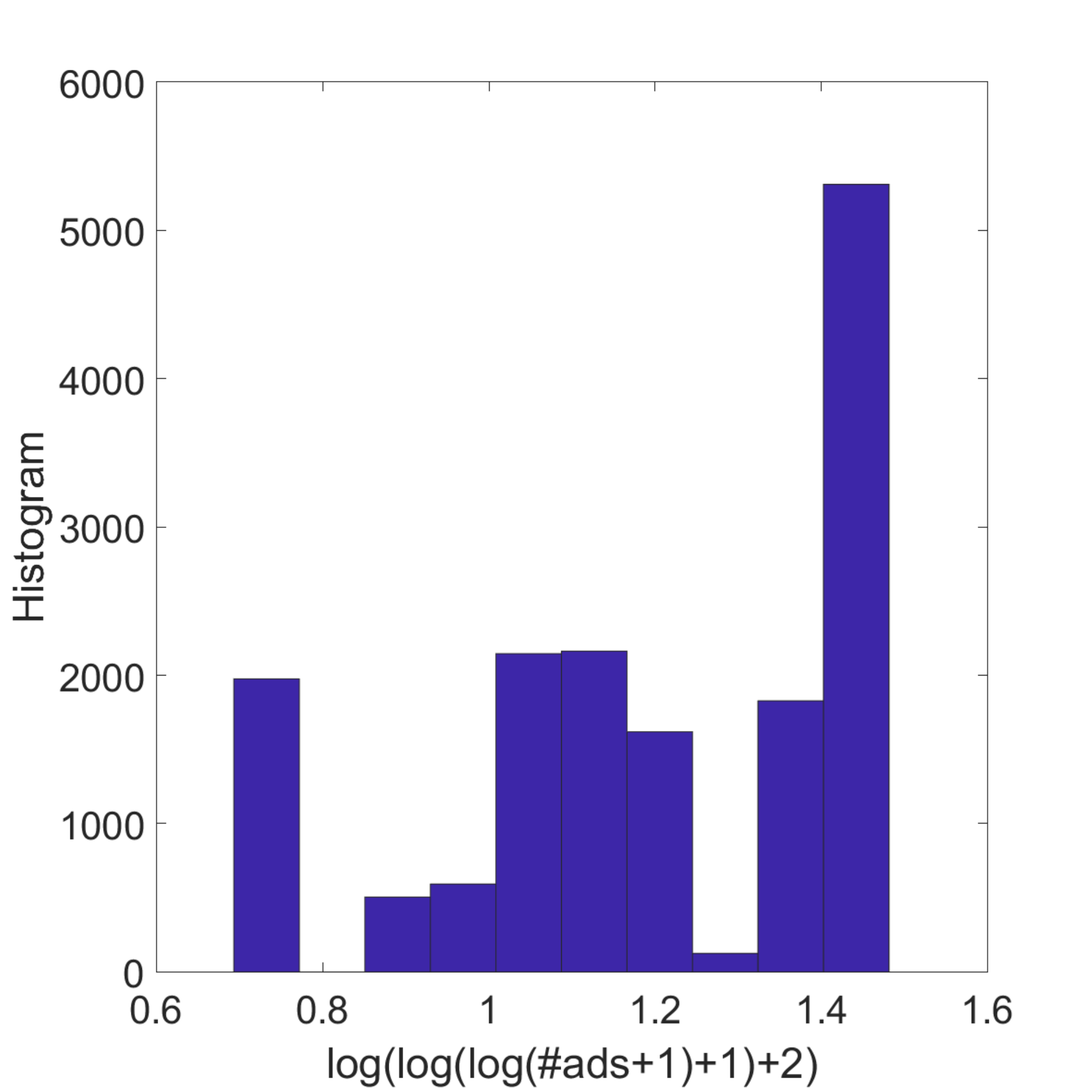}
			
			\caption{The histogram of the original campaign contribution data (top left) and the Box-Cox transformed contributions defined at \eqref{TransContributions} (bottom left), the histogram of the original counts of advertisements data (top right), and that of the transformed ones (bottom right).}\label{fig:HistScatter2}
		\end{figure}
		
		We reached the same conclusion in our data analysis.  Indeed, when we applied our method with logistic, cosine-sine, and indicator weight functions with a $B=500$ simulation-based approximation to test the model in \eqref{AiModel}, all the methods rejected the model with the $p$-values equal to 0.
		
		We examined the histogram of the original campaign contribution data and the number of advertisements $T$. From the first row of Figure~\ref{fig:HistScatter2}, we can see that both histograms are highly right-skewed. That is, they are not likely to fit any linear models. We then conducted a log-transformation, as in \cite{Ai_Linton_Motegi_Zhang_cts_treat}. However, the results were similar. 
		To make the data more likely to fit a linear model, we searched across Box-Cox transformations of the response data of the form $\text{BoxCox}(\text{Contribution},\lambda_1,\lambda_2):=\{(\text{Contribution}+\lambda_2)^{\lambda_1}-1\}/\lambda_1$ w.r.t. $\lambda_1, \lambda_2$ to find a transformation of the contribution whose sample quantiles have the largest correlation with those of a standard normal distribution. This yielded $(\tilde{\lambda}_1,\tilde{\lambda}_2) = (0.1397,0.0176)$. We then take 
		\begin{equation}
		Y = \text{BoxCox}(\text{Contribution},\tilde{\lambda}_1,\tilde{\lambda}_2)- \min\big\{\text{BoxCox}(\text{Contribution},\tilde{\lambda}_1,\tilde{\lambda}_2)\big\}\,,\label{TransContributions}
		\end{equation}
		so that the minimum response data is 0.
		The histogram of $Y$ is shown in the bottom left of Figure~\ref{fig:HistScatter2}. We can see that the transformed data remain highly right-skewed. However, now, it appears to be a truncated normal distribution.
		
		\begin{table}[H]
			\footnotesize
			\centering
			\caption{Estimated $p$-values from the U.S. presidential campaign data}\label{table:USpvalues}
			\resizebox{.6\textwidth}{!}{ 
				\begin{tabular}{cccc}
					\hline
					Statistic&  \text{Logistic}& \text{Cosine-Sine}& \text{Indicator} \\
					\hline
					CM& 0.610& 0.430& 0.740\\
					KS& 0.718& 0.667& 0.718\\
					\hline
				\end{tabular}
			}
		\end{table}
		
		\begin{figure}[H]
			\centering
			\includegraphics[width = .6\textwidth]{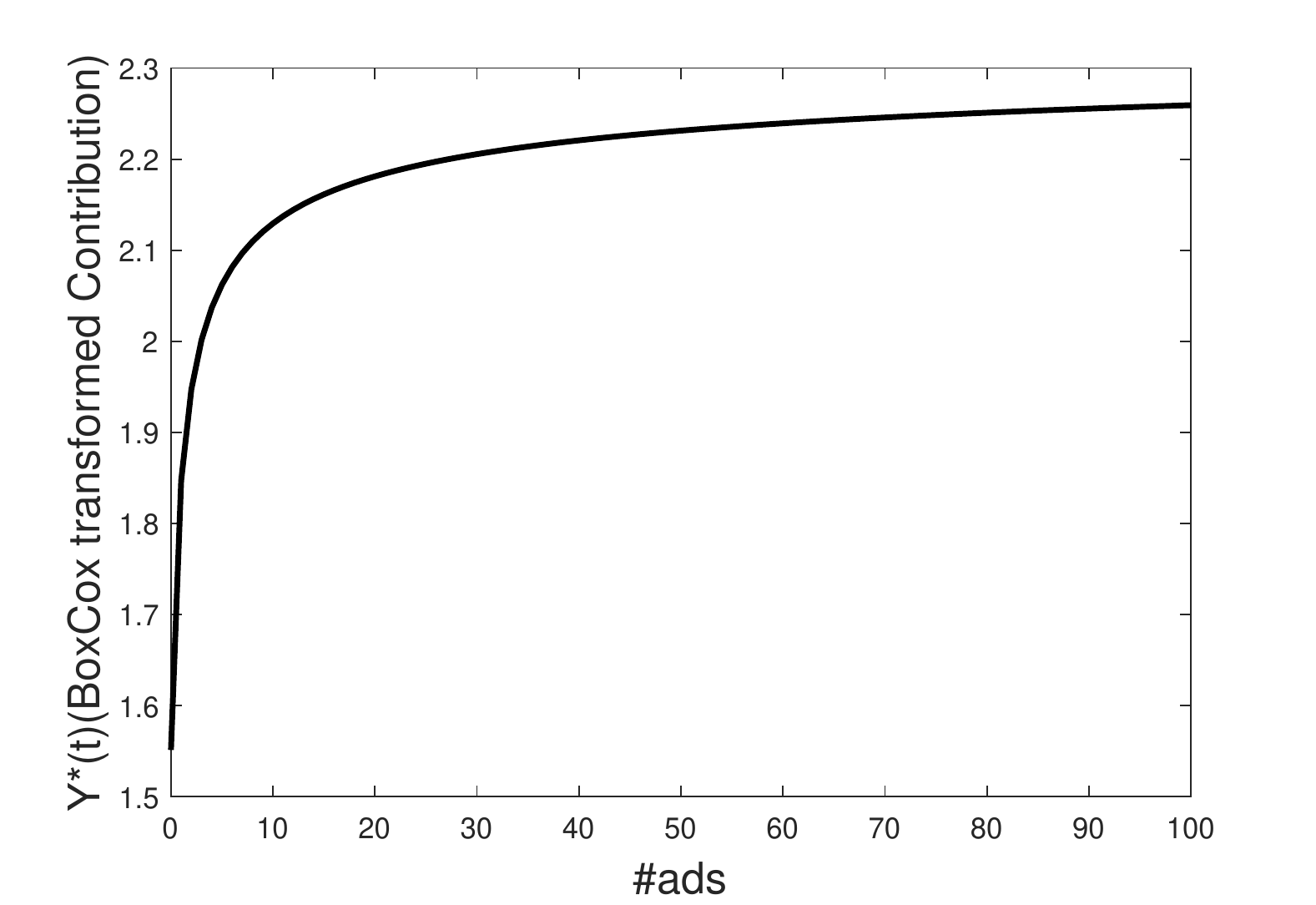}
			
			\caption{The plot of the estimated Tobit Model of the BoxCox transformed Campaign contribution versus the number of advertisements distributed.}\label{fig:EstModel}
		\end{figure}
		
		Now, it seems more reasonable to assume a Tobit model for the data. Specifically, let 
		$$
		Y(t) = \boldsymbol{\beta}^\top \boldsymbol{t} + \epsilon\,,
		$$
		for some unknown parameter $\boldsymbol{\beta}$ in a compact set in $\mathbb{R}^p$, where $\boldsymbol{t} = (1,t,t^2,\ldots,t^{p-1})^\top$ and $\epsilon$ is a normal random variable with mean 0 and unknown variance $\sigma^2$.
		We test a Tobit linear model on the potential outcome: 
		$$
		Y^\ast(t) = \begin{cases}
		Y(t) & \text{if} \quad Y(t)>0,\\
		0 & \text{if} \quad Y(t)\leq 0\,.
		\end{cases}
		$$ The details of the estimation of this model and the test statistics can be found in the supplementary material of this paper. We then tested the Tobit model with several different transformations of the treatment data \#ads and the polynomial order $p$. We found the Tobit model with $T = \log(\log(\log(\#\text{ads}+1)+1)+2)$ and $p=5$ gives the most reasonable results. The corresponding $p$-values are shown in Table~\ref{table:USpvalues}, and the estimated model is depicted in Figure~\ref{fig:EstModel}. It indicates that the campaign contribution increases rapidly with a relatively small increase in the number of advertisements from 0, and then the improvement gradually becomes marginal.

		\bigskip
		\begin{center}
			{\large\bf SUPPLEMENTARY MATERIAL}
		\end{center}
		
		Supplementary materials are only for online publication. The supplementary file contains  the simulation results of the KS-type statistic, the details of the estimation and the test statistics for the Tobit model used in section~\ref{sec:RealData}, the proofs of Theorems~\ref{Asymptotic:H0}, \ref{thm:asymp_pi0}, \ref{Asymptotic:H_L}, and the asymptotic properties of $\widehat{J}_N(t;\widehat{\boldsymbol{\theta}}_{opt})$ and $\widehat{CM}_N(\widehat{\boldsymbol{\theta}}_{opt})$.
		\if0\blind
		{
			\section*{Acknowledgments}
		We thank two  anonymous referees, an associate editor, and the editor Christian Hansen for constructive comments which lead to substantial improvement of the paper.	The first author, Wei Huang's research was supported by the Professor Maurice H. Belz Fund of the University of Melbourne.
			The second author, Oliver Linton, acknowledges Cambridge INET for financial support. 
			The final author, Zheng Zhang, acknowledges financial support from the National Natural Science Foundation of China through project 12001535, and the fund for building world-class universities (disciplines) of the Renmin University of China.
		} \fi
		
		\if1\blind
		{} \fi
		
		\bibliographystyle{Chicago}
		\bibliography{Semiparametric}

\begin{thebibliography}{}

\bibitem[\protect\citeauthoryear{Abadie and Cattaneo}{Abadie and
  Cattaneo}{2018}]{abadie2018econometric}
Abadie, A. and M.~D. Cattaneo (2018).
\newblock Econometric methods for program evaluation.
\newblock {\em Annual Review of Economics\/}~{\em 10}, 465--503.

\bibitem[\protect\citeauthoryear{Abrevaya, Hsu, and Lieli}{Abrevaya
  et~al.}{2015}]{abrevaya2015estimating}
Abrevaya, J., Y.-C. Hsu, and R.~P. Lieli (2015).
\newblock Estimating conditional average treatment effects.
\newblock {\em Journal of Business \& Economic Statistics\/}~{\em 33\/}(4),
  485--505.

\bibitem[\protect\citeauthoryear{Ai and Chen}{Ai and
  Chen}{2003}]{ai2003efficient}
Ai, C. and X.~Chen (2003).
\newblock Efficient estimation of models with conditional moment restrictions
  containing unknown functions.
\newblock {\em Econometrica\/}~{\em 71\/}(6), 1795--1843.

\bibitem[\protect\citeauthoryear{Ai, Huang, and Zhang}{Ai
  et~al.}{2020}]{ai2020mann}
Ai, C., L.~Huang, and Z.~Zhang (2020).
\newblock A mann--whitney test of distributional effects in a multivalued
  treatment.
\newblock {\em Journal of Statistical Planning and Inference\/}~{\em 209},
  85--100.

\bibitem[\protect\citeauthoryear{Ai, Huang, and Zhang}{Ai
  et~al.}{2022}]{aisimple}
Ai, C., L.~Huang, and Z.~Zhang (2022).
\newblock A simple and efficient estimation of average treatment effects in
  models with unmeasured confounders.
\newblock {\em Statistica Sinica\/}~{\em 32\/}(3).

\bibitem[\protect\citeauthoryear{Ai, Linton, Motegi, and Zhang}{Ai
  et~al.}{2021}]{Ai_Linton_Motegi_Zhang_cts_treat}
Ai, C., O.~Linton, K.~Motegi, and Z.~Zhang (2021).
\newblock A unified framework for efficient estimation of general treatment
  models.
\newblock {\em Quantitative Economics\/}~{\em 12\/}(3), 779--816.

\bibitem[\protect\citeauthoryear{Ai, Linton, and Zhang}{Ai
  et~al.}{2020}]{ai2020simple}
Ai, C., O.~Linton, and Z.~Zhang (2020).
\newblock A simple and efficient estimation method for models with nonignorable
  missing data.
\newblock {\em Statistica Sinica\/}~{\em 30}, 1949--1970.

\bibitem[\protect\citeauthoryear{Ai, Linton, and Zhang}{Ai
  et~al.}{2021}]{ai2021estimation}
Ai, C., O.~Linton, and Z.~Zhang (2021).
\newblock Estimation and inference for the counterfactual distribution and
  quantile functions in continuous treatment models.
\newblock {\em Journal of Econometrics\/}.

\bibitem[\protect\citeauthoryear{Ait-Sahalia, Bickel, and Stoker}{Ait-Sahalia
  et~al.}{2001}]{ait2001goodness}
Ait-Sahalia, Y., P.~J. Bickel, and T.~M. Stoker (2001).
\newblock Goodness-of-fit tests for kernel regression with an application to
  option implied volatilities.
\newblock {\em Journal of Econometrics\/}~{\em 105\/}(2), 363--412.

\bibitem[\protect\citeauthoryear{Andrews}{Andrews}{1994}]{andrews1994empirical}
Andrews, D. W.~K. (1994).
\newblock Empirical process methods in econometrics.
\newblock In R.~F. Engle and D.~L. McFadden (Eds.), {\em Handbook of
  Econometrics}, Volume~4, Chapter~37, pp.\  2247--2294. Citeseer.

\bibitem[\protect\citeauthoryear{Ao, Calonico, and Lee}{Ao
  et~al.}{2021}]{ao2021multivalued}
Ao, W., S.~Calonico, and Y.-Y. Lee (2021).
\newblock Multivalued treatments and decomposition analysis: An application to
  the wia program.
\newblock {\em Journal of Business \& Economic Statistics\/}~{\em 39\/}(1),
  358--371.

\bibitem[\protect\citeauthoryear{Athey, Imbens, and Wager}{Athey
  et~al.}{2018}]{Athey2018Approximate}
Athey, S., G.~W. Imbens, and S.~Wager (2018).
\newblock Approximate residual balancing: debiased inference of average
  treatment effects in high dimensions.
\newblock {\em Journal of the Royal Statistical Society: Series B (Statistical
  Methodology)\/}~{\em 80\/}(4), 597--623.

\bibitem[\protect\citeauthoryear{Bierens}{Bierens}{1982}]{bierens1982consistent}
Bierens, H.~J. (1982).
\newblock Consistent model specification tests.
\newblock {\em Journal of Econometrics\/}~{\em 20\/}(1), 105--134.

\bibitem[\protect\citeauthoryear{Bierens}{Bierens}{1990}]{bierens1990consistent}
Bierens, H.~J. (1990).
\newblock A consistent conditional moment test of functional form.
\newblock {\em Econometrica: Journal of the Econometric Society\/}, 1443--1458.

\bibitem[\protect\citeauthoryear{Bierens and Ploberger}{Bierens and
  Ploberger}{1997}]{bierens1997asymptotic}
Bierens, H.~J. and W.~Ploberger (1997).
\newblock Asymptotic theory of integrated conditional moment tests.
\newblock {\em Econometrica: Journal of the Econometric Society\/}, 1129--1151.

\bibitem[\protect\citeauthoryear{Blundell and Powell}{Blundell and
  Powell}{2003}]{blundell2001endogeneity}
Blundell, R. and J.~L. Powell (2003).
\newblock Endogeneity in nonparametric and semiparametric regression models.
\newblock ~{\em 8}.

\bibitem[\protect\citeauthoryear{Buchinsky}{Buchinsky}{1995}]{buchinsky1995quantile}
Buchinsky, M. (1995).
\newblock Quantile regression, box-cox transformation model, and the u.s. wage
  structure, 1963--1987.
\newblock {\em Journal of Econometrics\/}~{\em 65\/}(1), 109--154.

\bibitem[\protect\citeauthoryear{Cattaneo}{Cattaneo}{2010}]{cattaneo2010efficient}
Cattaneo, M.~D. (2010).
\newblock Efficient semiparametric estimation of multi-valued treatment effects
  under ignorability.
\newblock {\em Journal of Econometrics\/}~{\em 155\/}(2), 138--154.

\bibitem[\protect\citeauthoryear{Chan, Yam, and Zhang}{Chan
  et~al.}{2016}]{chan2016globally}
Chan, K. C.~G., S.~C.~P. Yam, and Z.~Zhang (2016).
\newblock Globally efficient non-parametric inference of average treatment
  effects by empirical balancing calibration weighting.
\newblock {\em Journal of the Royal Statistical Society: Series B (Statistical
  Methodology)\/}~{\em 78\/}(3), 673--700.

\bibitem[\protect\citeauthoryear{Chen}{Chen}{2007}]{chen2007large}
Chen, X. (2007).
\newblock Large sample sieve estimation of semi-nonparametric models.
\newblock {\em Handbook of Econometrics\/}~{\em 6\/}(B), 5549--5632.

\bibitem[\protect\citeauthoryear{Chen and Fan}{Chen and
  Fan}{1999}]{chen1999consistent}
Chen, X. and Y.~Fan (1999).
\newblock {Consistent hypothesis testing in semiparametric and nonparametric
  models for econometric time series.}
\newblock {\em Journal of Econometrics\/}~{\em 91}, 373--401.

\bibitem[\protect\citeauthoryear{Chen, Linton, and Van~Keilegom}{Chen
  et~al.}{2003}]{chen2003estimation}
Chen, X., O.~Linton, and I.~Van~Keilegom (2003).
\newblock Estimation of semiparametric models when the criterion function is
  not smooth.
\newblock {\em Econometrica\/}~{\em 71\/}(5), 1591--1608.

\bibitem[\protect\citeauthoryear{Chen, Hsu, and Wang}{Chen
  et~al.}{2020}]{chen2020stochastic}
Chen, Y.-T., Y.-C. Hsu, and H.-J. Wang (2020).
\newblock A stochastic frontier model with endogenous treatment status and
  mediator.
\newblock {\em Journal of Business \& Economic Statistics\/}~{\em 38\/}(2),
  243--256.

\bibitem[\protect\citeauthoryear{Chernozhukov, Fern{\'a}ndez-Val, and
  Melly}{Chernozhukov et~al.}{2013}]{chernozhukov2013inference}
Chernozhukov, V., I.~Fern{\'a}ndez-Val, and B.~Melly (2013).
\newblock Inference on counterfactual distributions.
\newblock {\em Econometrica\/}~{\em 81\/}(6), 2205--2268.

\bibitem[\protect\citeauthoryear{Colangelo and Lee}{Colangelo and
  Lee}{2020}]{colangelo2020double}
Colangelo, K. and Y.-Y. Lee (2020).
\newblock Double debiased machine learning nonparametric inference with
  continuous treatments.
\newblock {\em arXiv preprint arXiv:2004.03036\/}.

\bibitem[\protect\citeauthoryear{Crump, Hotz, Imbens, and Mitnik}{Crump
  et~al.}{2008}]{crump2008nonparametric}
Crump, R.~K., V.~J. Hotz, G.~W. Imbens, and O.~A. Mitnik (2008).
\newblock Nonparametric tests for treatment effect heterogeneity.
\newblock {\em The Review of Economics and Statistics\/}~{\em 90\/}(3),
  389--405.

\bibitem[\protect\citeauthoryear{de~Castro, Galvao, Kaplan, and Liu}{de~Castro
  et~al.}{2019}]{de2019smoothed}
de~Castro, L., A.~F. Galvao, D.~M. Kaplan, and X.~Liu (2019).
\newblock Smoothed gmm for quantile models.
\newblock {\em Journal of Econometrics\/}~{\em 213\/}(1), 121--144.

\bibitem[\protect\citeauthoryear{Donald and Hsu}{Donald and
  Hsu}{2014}]{Donald2014Estimation}
Donald, S.~G. and Y.-C. Hsu (2014).
\newblock Estimation and inference for distribution functions and quantile
  functions in treatment effect models.
\newblock {\em Journal of Econometrics\/}~{\em 178\/}(3), 383--397.

\bibitem[\protect\citeauthoryear{Donald, Hsu, and Lieli}{Donald
  et~al.}{2014}]{donald2014testing}
Donald, S.~G., Y.-C. Hsu, and R.~P. Lieli (2014).
\newblock Testing the unconfoundedness assumption via inverse probability
  weighted estimators of (l) att.
\newblock {\em Journal of Business \& Economic Statistics\/}~{\em 32\/}(3),
  395--415.

\bibitem[\protect\citeauthoryear{Dong, Lee, and Gou}{Dong
  et~al.}{2019}]{dong2019regression}
Dong, Y., Y.-Y. Lee, and M.~Gou (2019).
\newblock Regression discontinuity designs with a continuous treatment.
\newblock {\em Available at SSRN 3167541\/}.

\bibitem[\protect\citeauthoryear{Fan, Hsu, Lieli, and Zhang}{Fan
  et~al.}{2020}]{fan2020estimation}
Fan, Q., Y.-C. Hsu, R.~P. Lieli, and Y.~Zhang (2020).
\newblock Estimation of conditional average treatment effects with
  high-dimensional data.
\newblock {\em Journal of Business \& Economic Statistics\/}, 1--15.

\bibitem[\protect\citeauthoryear{Fan and Li}{Fan and
  Li}{1996}]{fan1996consistent}
Fan, Y. and Q.~Li (1996).
\newblock Consistent model specification tests: omitted variables and
  semiparametric functional forms.
\newblock {\em Econometrica: Journal of the econometric society\/}, 865--890.

\bibitem[\protect\citeauthoryear{Fan and Li}{Fan and
  Li}{2000}]{fan2000consistent}
Fan, Y. and Q.~Li (2000).
\newblock Consistent model specification tests: Kernel-based tests versus
  bierens' icm tests.
\newblock {\em Econometric Theory\/}, 1016--1041.

\bibitem[\protect\citeauthoryear{Firpo}{Firpo}{2007}]{Firpo2007Efficient}
Firpo, S. (2007).
\newblock Efficient semiparametric estimation of quantile treatment effects.
\newblock {\em Econometrica\/}~{\em 75\/}(1), 259--276.

\bibitem[\protect\citeauthoryear{Fong, Hazlett, and Imai}{Fong
  et~al.}{2018}]{Fong_Hazlett_Imai_2018}
Fong, C., C.~Hazlett, and K.~Imai (2018).
\newblock Covariate balancing propensity score for a continuous treatment:
  Application to the efficacy of political advertisements.
\newblock {\em Annals of Applied Statistics\/}~{\em 12\/}(1), 156--177.

\bibitem[\protect\citeauthoryear{Galvao and Wang}{Galvao and
  Wang}{2015}]{galvao2015uniformly}
Galvao, A.~F. and L.~Wang (2015).
\newblock Uniformly semiparametric efficient estimation of treatment effects
  with a continuous treatment.
\newblock {\em Journal of the American Statistical Association\/}~{\em
  110\/}(512), 1528--1542.

\bibitem[\protect\citeauthoryear{Hahn}{Hahn}{1998}]{hahn1998role}
Hahn, J. (1998).
\newblock {On the role of the propensity score in efficient semiparametric
  estimation of average treatment effects}.
\newblock {\em Econometrica\/}~{\em 66\/}(2), 315--331.

\bibitem[\protect\citeauthoryear{Hansen}{Hansen}{1982}]{hansen1982large}
Hansen, L. (1982).
\newblock Large sample properties of generalized method of moments estimators.
\newblock {\em Econometrica\/}~{\em 50}, 1029--1054.

\bibitem[\protect\citeauthoryear{Hirano and Imbens}{Hirano and
  Imbens}{2004}]{hirano2004propensity}
Hirano, K. and G.~W. Imbens (2004).
\newblock The propensity score with continuous treatments.
\newblock In A.~Gelman and X.-L. Meng (Eds.), {\em Applied Bayesian Modeling
  and Causal Inference from Incomplete-Data Perspectives}, Chapter~7, pp.\
  73--84. John Wiley \& Sons Ltd.

\bibitem[\protect\citeauthoryear{Hirano, Imbens, and Ridder}{Hirano
  et~al.}{2003}]{Hirano03}
Hirano, K., G.~W. Imbens, and G.~Ridder (2003, July).
\newblock Efficient estimation of average treatment effects using the estimated
  propensity score.
\newblock {\em Econometrica\/}~{\em 71\/}(4), 1161--1189.

\bibitem[\protect\citeauthoryear{Hsu, Lai, and Lieli}{Hsu
  et~al.}{2020}]{hsu2020counterfactual}
Hsu, Y.-C., T.-C. Lai, and R.~P. Lieli (2020).
\newblock Counterfactual treatment effects: Estimation and inference.
\newblock {\em Journal of Business \& Economic Statistics\/}, 1--16.

\bibitem[\protect\citeauthoryear{Huber, Hsu, Lee, and Lettry}{Huber
  et~al.}{2020}]{huber2020direct}
Huber, M., Y.-C. Hsu, Y.-Y. Lee, and L.~Lettry (2020).
\newblock Direct and indirect effects of continuous treatments based on
  generalized propensity score weighting.
\newblock {\em Journal of Applied Econometrics\/}~{\em 35\/}(7), 814--840.

\bibitem[\protect\citeauthoryear{Imai and Ratkovic}{Imai and
  Ratkovic}{2014}]{imai2014covariate}
Imai, K. and M.~Ratkovic (2014).
\newblock Covariate balancing propensity score.
\newblock {\em Journal of the Royal Statistical Society: Series B (Statistical
  Methodology)\/}~{\em 76\/}(1), 243--263.

\bibitem[\protect\citeauthoryear{Imai and van Dyk}{Imai and van
  Dyk}{2004}]{imai2004causal}
Imai, K. and D.~A. van Dyk (2004).
\newblock Causal inference with general treatment regimes: Generalizing the
  propensity score.
\newblock {\em Journal of the American Statistical Association\/}~{\em
  99\/}(467), 854--866.

\bibitem[\protect\citeauthoryear{Imbens and Newey}{Imbens and
  Newey}{2009}]{imbens2009identification}
Imbens, G.~W. and W.~K. Newey (2009).
\newblock Identification and estimation of triangular simultaneous equations
  models without additivity.
\newblock {\em Econometrica\/}~{\em 77\/}(5), 1481--1512.

\bibitem[\protect\citeauthoryear{Imbens and Wooldridge}{Imbens and
  Wooldridge}{2009}]{imbens2009recent}
Imbens, G.~W. and J.~M. Wooldridge (2009).
\newblock Recent developments in the econometrics of program evaluation.
\newblock {\em Journal of Economic Literature\/}~{\em 47\/}(1), 5--86.

\bibitem[\protect\citeauthoryear{Kennedy, Ma, McHugh, and Small}{Kennedy
  et~al.}{2017}]{kennedy2017non}
Kennedy, E.~H., Z.~Ma, M.~D. McHugh, and D.~S. Small (2017).
\newblock Non-parametric methods for doubly robust estimation of continuous
  treatment effects.
\newblock {\em Journal of the Royal Statistical Society: Series B (Statistical
  Methodology)\/}~{\em 79\/}(4), 1229--1245.

\bibitem[\protect\citeauthoryear{Lee}{Lee}{2018}]{lee2018efficient}
Lee, Y.-Y. (2018).
\newblock Efficient propensity score regression estimators of multivalued
  treatment effects for the treated.
\newblock {\em Journal of Econometrics\/}~{\em 204\/}(2), 207--222.

\bibitem[\protect\citeauthoryear{Li}{Li}{1999}]{li1999consistent}
Li, Q. (1999).
\newblock Consistent model specification tests for time series econometric
  models.
\newblock {\em Journal of Econometrics\/}~{\em 92\/}(1), 101--147.

\bibitem[\protect\citeauthoryear{Li, Hsiao, and Zinn}{Li
  et~al.}{2003}]{li2003consistent}
Li, Q., C.~Hsiao, and J.~Zinn (2003).
\newblock Consistent specification tests for semiparametric/nonparametric
  models based on series estimation methods.
\newblock {\em Journal of Econometrics\/}~{\em 112\/}(2), 295--325.

\bibitem[\protect\citeauthoryear{Newey}{Newey}{1997}]{Newey97}
Newey, W.~K. (1997, July).
\newblock Convergence rates and asymptotic normality for series estimators.
\newblock {\em Journal of Econometrics\/}~{\em 79\/}(1), 147--168.

\bibitem[\protect\citeauthoryear{Pakes and Pollard}{Pakes and
  Pollard}{1989}]{pakes1989simulation}
Pakes, A. and D.~Pollard (1989).
\newblock Simulation and the asymptotics of optimization estimators.
\newblock {\em Econometrica\/}~{\em 57\/}(5), 1027--1057.

\bibitem[\protect\citeauthoryear{Praestgaard and Wellner}{Praestgaard and
  Wellner}{1993}]{praestgaard1993exchangeably}
Praestgaard, J. and J.~A. Wellner (1993).
\newblock Exchangeably weighted bootstraps of the general empirical process.
\newblock {\em The Annals of Probability\/}~{\em 21\/}(4), 2053--2086.

\bibitem[\protect\citeauthoryear{Robins, Hern\'{a}n, and Brumback}{Robins
  et~al.}{2000}]{robins2000marginal}
Robins, J.~M., M.~A. Hern\'{a}n, and B.~Brumback (2000).
\newblock Marginal structural models and causal inference in epidemiology.
\newblock {\em Epidemiology\/}~{\em 11\/}(5), 550--560.

\bibitem[\protect\citeauthoryear{Rosenbaum and Rubin}{Rosenbaum and
  Rubin}{1983}]{rosenbaum1983central}
Rosenbaum, P.~R. and D.~B. Rubin (1983).
\newblock The central role of the propensity score in observational studies for
  causal effects.
\newblock {\em Biometrika\/}~{\em 70\/}(1), 41--55.

\bibitem[\protect\citeauthoryear{Rosenbaum and Rubin}{Rosenbaum and
  Rubin}{1984}]{rosenbaum1984reducing}
Rosenbaum, P.~R. and D.~B. Rubin (1984).
\newblock Reducing bias in observational studies using subclassification on the
  propensity score.
\newblock {\em J. Am. Statist. Ass.\/}~{\em 79\/}(387), 516--524.

\bibitem[\protect\citeauthoryear{Ruppert, Wand, and Carroll}{Ruppert
  et~al.}{2003}]{ruppert2003}
Ruppert, D., M.~P. Wand, and R.~J. Carroll (2003).
\newblock {\em Semiparametric Regression}.
\newblock Cambridge University Press.

\bibitem[\protect\citeauthoryear{Sant'Anna, Song, and Xu}{Sant'Anna
  et~al.}{2020}]{sant2018covariate}
Sant'Anna, P.~H., X.~Song, and Q.~Xu (2020).
\newblock Covariate distribution balance via propensity scores.
\newblock {\em arXiv preprint arXiv:1810.01370\/}.

\bibitem[\protect\citeauthoryear{Shaikh, Simonsen, Vytlacil, and Yildiz}{Shaikh
  et~al.}{2009}]{shaikh2009specification}
Shaikh, A.~M., M.~Simonsen, E.~J. Vytlacil, and N.~Yildiz (2009).
\newblock A specification test for the propensity score using its distribution
  conditional on participation.
\newblock {\em Journal of Econometrics\/}~{\em 151\/}(1), 33--46.

\bibitem[\protect\citeauthoryear{Stinchcombe and White}{Stinchcombe and
  White}{1998}]{stinchcombe1998consistent}
Stinchcombe, M.~B. and H.~White (1998).
\newblock Consistent specification testing with nuisance parameters present
  only under the alternative.
\newblock {\em Econometric theory\/}~{\em 14\/}(3), 295--325.

\bibitem[\protect\citeauthoryear{Stute}{Stute}{1997}]{stute1997nonparametric}
Stute, W. (1997).
\newblock Nonparametric model checks for regression.
\newblock {\em The Annals of Statistics\/}, 613--641.

\bibitem[\protect\citeauthoryear{Urban and Niebler}{Urban and
  Niebler}{2014}]{Urban_Niebler_2014}
Urban, C. and S.~Niebler (2014).
\newblock Dollars on the sidewalk: Should u.s. presidential candidates
  advertise in uncontested states?
\newblock {\em American Journal of Political Science\/}~{\em 58\/}(2),
  322--336.

\bibitem[\protect\citeauthoryear{Van Der~Vaart and Wellner}{Van Der~Vaart and
  Wellner}{1996}]{van1996weak}
Van Der~Vaart, A.~W. and J.~A. Wellner (1996).
\newblock {\em Weak convergence and empirical processes with applications to
  statistics}.
\newblock Springer.

\bibitem[\protect\citeauthoryear{Zheng}{Zheng}{1996}]{zheng1996consistent}
Zheng, J.~X. (1996).
\newblock A consistent test of functional form via nonparametric estimation
  techniques.
\newblock {\em Journal of Econometrics\/}~{\em 75\/}(2), 263--289.

\end{thebibliography}
		
	\end{document}